\documentclass[twocolumn,superscriptaddress,aps]{revtex4}
\usepackage{epsfig}
\usepackage{graphicx}
\usepackage{bm}

\begin{document}
\title{Quantum noise effects with Kerr nonlinearity enhancement in coupled gain-loss waveguides}
\author{Bing He}
\email{binghe@uark.edu}
\affiliation{Department of Physics, University of Arkansas, Fayetteville, Arkansas 72701, USA}
\author{Shu-Bin Yan}
\affiliation{Department of Physics, University of Arkansas, Fayetteville, Arkansas 72701, USA}
\affiliation{Science and Technology on Electronic Test \& Measurement Laboratory, North University of China, Taiyuan 030051, 
Shanxi, China}
\author{Jing Wang}
\affiliation{Department of Physics, University of Arkansas, Fayetteville, Arkansas 72701, USA}
\author{Min Xiao}
\email{mxiao@uark.edu}
\affiliation{Department of Physics, University of Arkansas, Fayetteville, Arkansas 72701, USA}
\affiliation{National Laboratory of Solid State Microstructures and School of Physics, Nanjing University, Nanjing 210093, 
China}

\begin{abstract}
It is generally difficult to study the dynamical properties of a quantum system with both inherent quantum noises and 
non-perturbative nonlinearity. Due to the possibly drastic intensity increase of an input coherent light in the gain-loss 
waveguide couplers with parity-time ($\mathcal{PT}$) symmetry, the Kerr effect from a nonlinearity added into the systems can 
be greatly enhanced, and is expected to create the macroscopic entangled states of the output light fields with huge photon numbers. 
Meanwhile, the quantum noises also coexist with the amplification and dissipation of the light fields. Under the interplay between 
the quantum noises and nonlinearity, the quantum dynamical behaviors of the systems become rather complicated. However, the important quantum noise effects have been mostly neglected in the previous studies about nonlinear $\mathcal{PT}$-symmetric systems. Here we present a solution to this non-perturbative quantum nonlinear problem, showing the real-time evolution of the system observables. The enhanced Kerr nonlinearity is found to give rise to a previously unknown decoherence effect that is irrelevant to the quantum noises, and imposes a limit on the emergence of macroscopic nonclassicality. In contrast to what happen in the linear systems, the quantum noises exert significant impact on the system dynamics, and can create the nonclassical light field states in conjunction with the enhanced Kerr nonlinearity. This first study on the noise involved quantum nonlinear dynamics of the coupled gain-loss waveguides 
can help to better understand the quantum noise effects in the broad nonlinear systems.
\end{abstract}
\maketitle

\section{introduction}
Originating from the uncertainty relation in quantum mechanics, quantum noises are ubiquitous in open quantum systems coupled to their environment. Those accompanying light dissipation are unavoidable in any realistic quantum optical system \cite{book}, while the noise going together with light amplification determines the quantum limit of an amplifier \cite{rv}. These most commonly encountered quantum noises can be regarded as the random drives from the associated external reservoirs interacting with a quantum system, and they also follow the law of quantum mechanics. The solvability of linear Heisenberg-Langevin equations makes it possible to describe the quantum noise effects in systems with quadratic Hamiltonians, but the situations of nonlinear quantum systems are much more complicated. 
In classical nonlinear systems, noises as the fluctuations of system parameters or random external forces are known to give rise to 
various interesting physical effects such as stochastic resonance \cite{sr}, noise-induced phase transition \cite{n-p-t} and phase synchronization \cite{ns}, etc. However, the effects of quantum noises, especially those in the systems with non-perturbative nonlinearity, remained to be uncovered yet.

Open quantum systems with Kerr nonlinearity are meaningful examples for studying quantum noise effects. Kerr nonlinearity is considered as a prerequisite for generating macroscopic photonic states \cite{y-s} and operating deterministic quantum logic devices \cite{l-o}. The Kerr coefficient in a natural material is typically small, though it can be enhanced in coherently prepared atomic ensembles 
\cite{wang} or Josephson junctions \cite{j1,j2}. The straightforward way to get a larger Kerr nonlinear effect on an input light is to strengthen its intensity. Such seemingly trivial practice of enhancing Kerr nonlinearity can lead to interesting phenomena in a simple system as illustrated in Fig. 1. Here two coupled waveguides are with the balanced gain and loss rates, respectively. This model has attracted intensive researches in recent years as it realizes an optical analogue of parity-time ($\mathcal{PT}$)-symmetric quantum mechanics \cite{bender, bender2}, and some recent experiments with similar systems \cite{ex1, ex2, ex3, ex4} have demonstrated its interesting light transmission properties. In such systems the light intensity undergoes a drastic transition from periodic oscillation to exponential increase when they are tuned into the regime of $\mathcal{PT}$ symmetry broken. If one of the waveguides is also added with a weak Kerr nonlinearity, one will see its significant influence on the light field dynamics because its effects are greatly enhanced to the non-perturbative ones in the symmetry broken regime. In a Kerr medium without gain or loss, an input coherent light will evolve into a so-called Kerr state that might manifest macroscopic nonclassicality (see, e.g. the review articles \cite{kerr1, jeong}), and its evolution to a photonic Schr\"{o}dinger cat state (the superposition of two coherent states with the equal but opposite sign of amplitudes) of a few photons has been demonstrated with a circuit QED setup that realizes an effective strong Kerr nonlinearity \cite{cat}. By making use of the above mentioned Kerr effect enhancement from light amplification, two-mode macroscopic nonclassical 
states such as entangled states of light fields, are expected to be generated. On the other hand, the existing quantum noises with the gain and loss of the light fields could destroy such nonclassicality as in other quantum systems \cite{esd}. To clarify these possibilities, it is necessary to find a complete solution to the system dynamics including the noise effects.

\begin{figure}[t!]
\vspace{0cm}
\centering
\epsfig{file=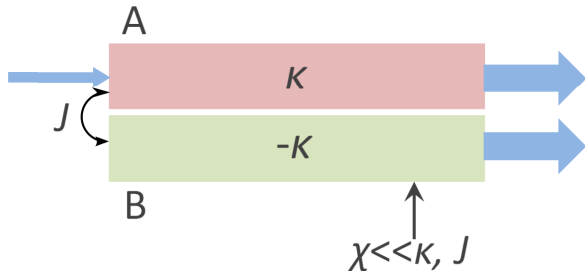,width=0.89\linewidth,clip=} 
{\vspace{-0.3cm}\caption{(color online) Setup. The light is amplified at the rate $\kappa$ in channel $A$, but damps at the same rate in channel $B$. 
The two wave-guide modes couple at the rate $J$. The loss channel also carries a Kerr nonlinearity with small coefficient $\chi$.}}
\vspace{0.2cm}
\end{figure}

The similar setups with the assumed strong nonlinearity \cite{n1,n2,n3, n5,n7,n8,n9} were theoretically studied in the context 
of all-optical signal control such as non-reciprocal light propagation, and other Kerr type classical nonlinear 
$\mathcal{PT}$-symmetric systems were also explored (see, e.g. \cite{no1,no2,no3,no4,no5, no6}). In addition to the above researches on the classical aspect of nonlinear $\mathcal{PT}$-symmetric systems, the quantum properties of some $\mathcal{PT}$-symmetric nonlinear systems have received attentions recently \cite{jing, da}. 
However, except for the noise-induced spontaneous photon generation in linear couplers \cite{arga} and their generalized structures 
\cite{CPT}, little was known about the effects of the quantum noises in these $\mathcal{PT}$-symmetric nonlinear systems. An obvious difficulty in approaching the dynamics of these systems is the non-integrability of their dynamical equations in the presence of the noise terms. Since the significant enhancement of Kerr effect is from the exponentially growing light intensity after breaking the $\mathcal{PT}$ symmetry, there is no steady-state solution in the course of evolution to the non-perturbative nonlinear regime. So far the linearization of the dynamical equations 
around its steady-state solutions or other mean values of the associated system operators is the most commonly adopted approach to a general quantum nonlinear system \cite{hillery} (the beyond linearization approach for finding the steady-state quasi-probability function of a special nonlinear system can be seen in \cite{d-w}), but it is not workable for studying the dynamical process in the systems. Generally, few approaches except for numerical simulation \cite{num} have been reported for dealing with the dynamics of a nonlinear system in non-perturbative regime and under quantum noise effects at the same time. 

Here, we present a first solution to such challenging dynamical problem for a coherent light sent into the nonlinear coupler in Fig. 1. A novel phenomenon found by our approach is the decoherence of an input coherent light going together with the Kerr nonlinearity enhancement in the symmetry breaking regime, but it does not originate from the quantum noise effects as in other quantum systems. On the other hand, the quantum noises significantly influence the dynamics of this nonlinear system, deviating the system observable evolutions from those predicted by the non-Hermitian effective Hamiltonian neglecting the quantum noises. Moreover, analogous to 
the synergic effects in the classical nonlinear systems \cite{sr, n-p-t, ns}, the joint action of the quantum noises and enhanced nonlinearity widens the regimes for the existence of the macroscopic nonclassical states.   

\section{Quantum Dynamics of coupled gain-loss waveguides} 

Under the full dynamics including the light field coupling under the balanced gain and loss, as well as the nonlinearity of the Kerr coefficient $\chi$, the Heisenberg-Langevin equations for the two waveguide modes in Fig. 1 (with the notation $\hbar\equiv 1$) read 
\begin{eqnarray}
i\frac{d}{dt}\hat{a}(t)&=&i\kappa\hat{a}(t)+J\hat{b}(t)+i\sqrt{2\kappa}\hat{\xi}^\dagger_a(t),\nonumber\\
i\frac{d}{dt}\hat{b}(t)&=&-i\kappa\hat{b}(t)+J\hat{a}(t)+\chi\hat{b}^\dagger(t)\hat{b}(t)\hat{b}(t)\nonumber\\
&+&i\sqrt{2\kappa}\hat{\xi}_b(t).
\label{dynamics}
\end{eqnarray}
Here the quantum noise $\hat{\xi}^\dagger_a(\hat{\xi}_b)$ coexisting with the light field amplification (dissipation) at the rate $\kappa$ 
also acts on the waveguide mode coupling at the rate $J$ with the other one, and the noise operator satisfies the commutation relation $[\hat{\xi}_c(t),\hat{\xi}_c^\dagger(t')]=\delta(t-t')$ and the correlations $\langle \hat{\xi}^\dagger_c(t)\hat{\xi}_c(t')\rangle=0$, $\langle \hat{\xi}_c(t)\hat{\xi}_c^\dagger(t')\rangle=\delta(t-t')$ for $c=a, b$. Note that the notation for the amplification noise here is different from that in \cite{book, arga}, so that the correlations for both amplification and dissipation noise can be written in the unified forms. The input light we work with is in the coherent state $|\alpha_0\rangle$, and initially enters the gain channel without loss of generality. We assume $|\alpha_0|\gg 1$ 
with an aim to create macroscopic nonclassical states, but the product $\chi|\alpha_0|^2$ is small so that the Kerr effect is still weak from the input. 

The properties of the dynamical process in (\ref{dynamics}) can be seen better from the attempted approaches to the problem by the standard methods. The similar equations were discussed in a mean field approach which replaces the mode operator $\hat{a}$ ($\hat{b}$) with its expectation value $\alpha=\langle\hat{a}\rangle$ ($\beta=\langle\hat{b}\rangle$) \cite{n1,n2,n3, n5, n7,n8,n9,no1,no2,no3,no4,no5,no6} and has the noise terms averaged out, i.e.
\begin{eqnarray}
i\frac{d}{dt}\alpha(t)&=&i\kappa\alpha(t)+J\beta(t),\nonumber\\
i\frac{d}{dt}\beta(t)&=&-i\kappa\beta(t)+J\alpha(t)+\chi|\beta(t)|^2\beta(t).
\label{c-dynamics}
\end{eqnarray}
Though the similar nonlinear differential equations in this mean field approach can be reduced to the solvable ones (see, e.g. \cite{n1, no5}), the above dynamical equations are not equivalent to the averaged quantum dynamical equations from (\ref{dynamics}). Obviously the mean value $\langle \hat{b}^\dagger\hat{b}\hat{b}\rangle$ is not exactly equal to $|\beta|^2\beta$, considering the quantum noise effects such as spontaneous photon generation \cite{arga}. Below we will also see other difference between the averages $\langle \hat{b}^\dagger\hat{b}\rangle$ and $\langle \hat{b}^\dagger\rangle\langle\hat{b}\rangle$ for this quantum nonlinear system. 

Another straightforward solution following the practice of numerical simulation \cite{num} gives the iterative forms of the evolved modes 
\begin{eqnarray}
\hat{a}[n+1]&=& \big(\sum_{c=a,b}M_{a,c}\hat{c}[n]+\sum_{\beta=\zeta_a^\dagger,\zeta_b}N_{a,\beta}\hat{\beta}[n]\big)\delta t,\nonumber\\
\hat{b}[n+1]&=& \big(\sum_{c=a,b}M_{b,c}\hat{c}[n]+\sum_{\beta=\zeta_a^\dagger,\zeta_b}N_{b,\beta}\hat{\beta}[n]\big)\delta t \nonumber\\
&+& e^{-i\chi\hat{b}^\dagger[n]\hat{b}[n]\delta t}\hat{b}[n]
\label{step}
\end{eqnarray}
at each step of evolution from $n\delta t$ to $(n+1)\delta t$, in which the matrix $\hat{M}$ is from the linear coupling of the waveguide modes plus the amplification and dissipation of the light fields, and the matrix $\hat{N}$ is due to the noise drives. For this quantum system, however, the linear coupling between the modes, as well as to the noises, and the action of the Kerr nonlinearity are not commutative. The above procedure of their independent actions is therefore consistent with the real system evolution only in the limit of infinite number of iterative steps ($\delta t\rightarrow 0$). 
The errors from the non-commutativity of the linear and nonlinear actions will accumulate for any finite $\delta t$.   

One could also take an alternative route in the Schr\"{o}dinger picture. The commonly used tool is the quantum master equation 
\begin{eqnarray}
\dot{\rho}=-i[J(\hat{a}\hat{b}^\dagger+\hat{a}^\dagger\hat{b})+\frac{1}{2}\chi(\hat{b}^\dagger)^2\hat{b}^2,\rho]+{\cal L}_{a}\rho+{\cal L}_{b}\rho
\label{m-e}
\end{eqnarray}
for the density matrix $\rho$ of the system,
where the super-operator operation in the Lindblad form
\begin{eqnarray}
{\cal L}_{a}\rho=-\kappa(\hat{a}\hat{a}^\dagger \rho+\rho\hat{a}\hat{a}^\dagger-2\hat{a}^\dagger\rho\hat{a})
\end{eqnarray}
describes the amplification process, and the corresponding one
\begin{eqnarray}
{\cal L}_{b}\rho=-\kappa(\hat{b}^\dagger\hat{b} \rho+\rho\hat{b}^\dagger\hat{b}-2\hat{b}\rho\hat{b}^\dagger)
\end{eqnarray}
is about the dissipation process. 
The analytical solutions to the master equations with Kerr nonlinearity were found only for the single mode situation thus far 
(see, e.g. \cite{m1,m2,m3}). In Ref. \cite{da}, for example, the dynamics described by a similar master equation is studied via 
a conversion to that of the mean values of a few system operators in the above mentioned mean field approach. Moreover, because the master equation is obtained by averaging out the noise-reservoir part 
in the momentary dynamical evolution of a joint quantum state of the system plus reservoirs (see Appendix A), it is difficult to 
see the involved quantum noise effects as in the mean field approach to the Heisenberg-Langevin equations.   

\section{Transition from perturbative to non-perturbative nonlinear dynamics} 

Our approach to this dynamical problem is based on the observation of the following Kerr nonlinearity transition 
across the threshold of $\mathcal{PT}$ symmetry. One sees the jump of the light intensity from the solution
\begin{eqnarray}
&&\left(
\begin{array}
[c]{c}
\hat{A}(t)\\
\hat{B}(t)
\end{array}
\right)=\left(
\begin{array}
[c]{c}
-ie^{-\lambda t}\frac{\eta_1}{J}\hat{o}_1+i
e^{\lambda t}\frac{\eta_2}{J}\hat{o}_2\\
e^{-\lambda t}\hat{o}_1+e^{\lambda t}\hat{o}_2
\end{array}
\right)\nonumber\\
&+&\sqrt{2\kappa} 
\int_0^t d\tau\left(
\begin{array}
[c]{c}
-ie^{-\lambda (t-\tau)}\frac{\eta_1}{J}\hat{n}_1(\tau)+i
e^{\lambda (t-\tau)}\frac{\eta_2}{J}\hat{n}_2(\tau)\\
e^{-\lambda (t-\tau)}\hat{n}_1(\tau)+e^{\lambda (t-\tau)}\hat{n}_2(\tau)
\end{array}
\right)\nonumber\\
\label{l-solution}
\end{eqnarray}
to (\ref{dynamics}) in the linear coupler limit with $\chi=0$, 
where $\hat{o}(\hat{n})_1=i\frac{J}{\eta_1+\eta_2}\hat{a}(\hat{\xi}_a^\dagger)+\frac{\eta_2}{\eta_1+\eta_2}\hat{b}(\hat{\xi}_b)$ 
and $\hat{o}(\hat{n})_2=-i\frac{J}{\eta_1+\eta_2}\hat{a}(\hat{\xi}_a^\dagger)+\frac{\eta_1}{\eta_1+\eta_2}\hat{b}(\hat{\xi}_b)$ with $\eta_{1(2)}=\mp\kappa+ \sqrt{\kappa^2-J^2}$ and $\lambda=\sqrt{\kappa^2-J^2}$. The linear coupler mode 
$\hat{A}(\hat{B})$ oscillates in the $\mathcal{PT}$-symmetric regime of $\kappa< J$, but will exponentially grow if the symmetry is broken when $\kappa> J$. Given a weak Kerr nonlinearity as in Fig. 1, such drastic increase of the light intensity across the threshold $\kappa=J$ can greatly enhance its effects. Meanwhile, one noise component in (\ref{l-solution}) also becomes much larger in the symmetry broken regime, contributing to a more significant spontaneous photon generation.

Then we reformulate the process in Eq. (\ref{dynamics}) in term of the stochastic Hamiltonian
\begin{eqnarray}
H_{L}(t)&=&J(\hat{a}\hat{b}^\dagger+\hat{a}^\dagger\hat{b})+i\sqrt{2\kappa}\{\hat{a}^\dagger \hat{\xi}^\dagger_a(t)-\hat{a} \hat{\xi}_a(t)\}\nonumber\\
&+&i\sqrt{2\kappa}\{\hat{b}^\dagger \hat{\xi}_b(t)-\hat{b} \hat{\xi}^\dagger_b(t)\}, 
\label{H}
\end{eqnarray}
which does not commute with the additional Kerr nonlinear term $H_{NL}=1/2\chi(\hat{b}^\dagger)^2\hat{b}^2$. The construction of this stochastic Hamiltonian is discussed in Appendix A. The system evolves under the joint operation $U_L(t)$ as a time-ordered exponential $\mathcal{T}e^{-i\int_0^t d\tau H_L(\tau)}$ on both system and reservoirs, together with a perturbation of $H_{NL}$ when its effect is weak. Different from the last term involving the dissipation noise in (\ref{H}), a vacuum state will not be kept invariant if acting the second term of the squeezing type on it. This property explains the phenomenon of spontaneous photon generation. With this property an input coherent state $|\alpha_0\rangle_a$ will not simply evolve to an amplified coherent state $|e^{\lambda t}\alpha_0\rangle_a$ in the limit of no waveguide coupling ($J\rightarrow 0$).

To solve the dynamics of the quantum nonlinear system, we expand its evolution operator in terms of the perturbative Hamiltonian $H_{NL}$ as follows:
\begin{eqnarray}
&&U(t)=\mathcal{T}e^{-i\int_0^t d\tau (H_L+H_{NL})(\tau)}\nonumber\\
&=& U_L(t)
\big\{I-i\int_0^t ds_1 U_L^\dagger (s_1)H_{NL}U_L(s_1)
-\int_0^t ds_1 U_L^\dagger (s_1)\nonumber\\
&\times & H_{NL} U_L(s_1)\int_0^{s_1}ds_2 U_L^\dagger (s_2)H_{NL}U_L(s_2)
+\cdots \big\}.
\label{fact}
\end{eqnarray}
The proof of this expansion is given in Appendix B. This series expansion can also be written as a time-ordered exponential $\mathcal{T}e^{-i\int_0^t d\tau U^\dagger_L(\tau)H_{NL}U_L(\tau)}$ denoted as $U_{NL}(t)$. Similar methods of factorizing the actions of
different Hamiltonians in other physical systems can be found in \cite{U1, U2, U3}.
As seen from (\ref{l-solution}), the transformed Hamiltonian $U^\dagger_L(\tau)H_{NL}U_L(\tau)=1/2\chi(\hat{B}^\dagger(\tau))^2(\hat{B}(\tau))^2$ by the linear action $U_L(\tau)$ takes a transition between a perturbative and a non-perturbative term across the threshold $\kappa=J$, thus reflecting the physics of the Kerr nonlinearity in the system. 

\section{Solution to dynamically evolved waveguide modes} 

Using the infinite product form $\prod_i e^{-i U^\dagger_L(t_i)H_{NL}U_L(t_i)\delta t}$ of the nonlinear action $U_{NL}(t)$, one will find the exact forms of the evolved modes with Eq. (\ref{fact}) as follows:
\begin{eqnarray}
&&U^\dagger(t)\hat{a}U(t)=U_{NL}^\dagger(t)\hat{A}(t)U_{NL}(t)=\hat{A}(t)\nonumber\\
&-&i\chi\int_0^t d\tau c_{ab}(t,\tau)U_{NL}^\dagger(\tau)\hat{B}^\dagger(\tau)\hat{B}(\tau)\hat{B}(\tau)U_{NL}(\tau),\nonumber\\
&&U^\dagger(t)\hat{b}U(t)=U_{NL}^\dagger(t)\hat{B}(t)U_{NL}(t)=\hat{B}(t)\nonumber\\
&-&i\chi\int_0^t d\tau c_{bb}(t,\tau)U_{NL}^\dagger(\tau)\hat{B}^\dagger(\tau)\hat{B}(\tau)\hat{B}(\tau)U_{NL}(\tau),\nonumber\\
\label{exact}
\end{eqnarray}
where $c_{ab}(t,t')=[\hat{A}(t),\hat{B}^\dagger(t')]$, $c_{bb}(t,t')=[\hat{B}(t),\hat{B}^\dagger(t')]$ are the commutators of the linear coupler modes in (\ref{l-solution}). The procedure to obtain these evolved modes is similar to that in Eq. (C-4) of Appendix C. Iteratively applying the transformation
\begin{eqnarray}
&&U_{NL}^\dagger(\tau)\hat{B}(\tau)U_{NL}(\tau)=\hat{B}(\tau)\nonumber\\
&-&i\chi\int^{\tau}_0 dt' c_{bb}(\tau, t')U_{NL}^\dagger(t')\hat{B}^\dagger(t')\hat{B}(t')\hat{B}(t')U_{NL}(t')\nonumber\\
\end{eqnarray}
in Eq. (\ref{exact}) leads to two series expansions of the folded integrals to all orders of the Kerr coefficient $\chi$. 
These general forms of the evolved modes are valid in any regime of the system parameters. 

In the symmetry broken regime, the evolved modes 
can be reduced to 
\begin{eqnarray}
&&U^\dagger(t)\hat{a}U(t)= i \frac{\eta_2}{J}\mathcal{T}e^{i\zeta_2\chi\int_0^t d\tau \hat{B}^\dagger\hat{B}(\tau)}\hat{B}(t)\nonumber\\
&+& i\frac{\eta_2}{J}\sum_{n=1}^\infty(i\zeta_2\chi)^n\int_0^t dt_1 \hat{B}^\dagger\hat{B}(t_1)\int_0^{t_1} dt_2 \hat{B}^\dagger\hat{B}(t_2)\cdots\nonumber\\
&\times & \int_0^{t_{n-1}}dt_n \hat{B}^\dagger\hat{B}(t_n)\hat{\nu}(t,t_{n}),\nonumber\\
&&U^\dagger(t)\hat{b}U(t) =  \mathcal{T} e^{i\zeta_2\chi\int_0^t d\tau \hat{B}^\dagger\hat{B}(\tau)}\hat{B}(t)\nonumber\\
&+& \sum_{n=1}^\infty(i\zeta_2\chi)^n\int_0^t dt_1 \hat{B}^\dagger\hat{B}(t_1)\int_0^{t_{1}}dt_2 \hat{B}^\dagger\hat{B}(t_2)\cdots\nonumber\\
&\times & \int_0^{t_{n-1}}dt_n \hat{B}^\dagger\hat{B}(t_n)\hat{\nu}(t,t_{n}),
\label{solution}
\end{eqnarray}
where $\hat{\nu}(t,t_n)=\sqrt{2\kappa}\int_t^{t_n}d\tau e^{\lambda(t-\tau)}\hat{n}_2(\tau)$ is a noise operator. The derivation of the result is given in Appendix C. The coefficient $\zeta_2=\frac{\kappa}{\lambda}\frac{\eta_1^2-J^2}{(\eta_1+\eta_2)^2}$ is purely due to the quantum noise operator $\hat{n}_{2}$ in (\ref{l-solution}), which includes both amplification and dissipation noise. The reservoir degrees of freedom are therefore crucial to the evolved waveguide modes. Another interesting feature of this non-perturbative solution is that the mode $U^\dagger(t)\hat{a}U(t)$ out of the gain channel happens to be $i\eta_2/J$ times of the mode $U^\dagger(t)\hat{b}U(t)$ out of the loss channel, due to the forms of the evolved linear coupler modes in Eq. (\ref{l-solution}). 

To illustrate the quantum dynamics of the system, one should find the expectation values of the quantum operators evolving under the full dynamics including the quantum noise effects. For our input in the continuous-variable (CV) state $\rho_{in}=|\alpha_0\rangle_a\langle \alpha_0|$, a suitable operator that coveys much information is the quadrature $\hat{X}_c(\phi)=1/2(\hat{c}e^{-i\phi}+\hat{c}^\dagger e^{i\phi})$ of the evolving field modes $\hat{c}(t)=\hat{a}(t)$ and $\hat{b}(t)$. Going back to the Schr\"{o}dinger picture, the non-zero expectation of a quadrature means the existence of the quantum coherence from the superposition of the Fock basis, i.e. 
the presence of the off-diagonal elements $|n\rangle\langle n+1|$ and $|n+1\rangle\langle n|$ ($n\geq 0$) of the evolved density matrix. We apply the mean quadrature values in the different direction  $\phi$ to test if the initial coherent state could evolve to a less coherent one. In finding its expectation value $\langle \hat{X}_c(\phi)\rangle$ we need to average over both system and reservoir degrees of freedom. It can be done by first tracing out the reservoir part as in the following procedure:
\begin{eqnarray}
&&\langle \hat{b}(t)\rangle=\mbox{Tr}_{S,R} \big \{U^\dagger(t)\hat{b}U(t) \times |\alpha_0\rangle_a\langle \alpha_0|\otimes \rho_{R}\big\}\nonumber\\
&=& _a\big \langle \alpha_0 \big|e^{i\zeta_2\chi\int_0^t d\tau \{e^{2\lambda \tau}\hat{o}_2^\dagger\hat{o}_2+\sigma(\tau)\}}e^{\lambda t}\hat{o}_2\big|\alpha_0 \big\rangle_a\nonumber\\
&+ &_a\big \langle \alpha_0 \big| e^{i\zeta_2\chi \int_0^t dt' \{e^{2\lambda t'}\hat{o}_2^\dagger\hat{o}_2+\sigma(t')\}}i\zeta_2\chi \int_0^t d\tau \sigma(\tau) \nonumber\\
&\times &e^{i\zeta_1\zeta_2\chi\int^\tau_0 e^{2\lambda t'}dt'} e^{\lambda t}\hat{o}_2\big|\alpha_0 \big\rangle_a\nonumber\\
&\equiv & EB_0(t)+EB_1(t),
\label{average}
\end{eqnarray}
where 
\begin{eqnarray}
\sigma (\tau)=\frac{\kappa}{\lambda}(\frac{J}{\eta_1+\eta_2})^2(e^{2\lambda \tau}-1),
\end{eqnarray}
and $\zeta_1=[\hat{o}_2,\hat{o}_2^\dagger]=\frac{\eta_1^2+J^2}{(\eta_1+\eta_2)^2}$.
We use Wick's theorem to obtain the result in (\ref{average}).
The function $\sigma(t)$ in the phase factor of the first average $EB_0(t)$ is obtained by averaging over the noise operators inside 
the non-Abelian phase factor $\mathcal{T}e^{i\zeta_2\chi\int_0^t d\tau \hat{B}^\dagger\hat{B}(\tau)}$ in (\ref{solution}) over the reservoir state $\rho_{R}$. The operator inside this non-Abelian phase factor becomes commutative at the different time after averaging out the noise part, and then it is reduced to an ordinary exponential. The integral in the second term $EB_1(t)$ is found by averaging the noise part in the operator $\hat{B}(t)$ on the right side of (\ref{solution}) with its counterpart in the above mention phase operator. The terms containing the noise operator 
$\hat{\nu}(t,t_n)$ in (\ref{solution}) make negligible contribution to the expectation value given a strong input beam. 

As a comparison, we also consider the waveguide mode evolution under the effective Hamiltonian $H_{PT}+H_{NL}$ 
without quantum noises. Here the non-Hermitian $\mathcal{PT}$-symmetric Hamiltonian
\begin{eqnarray}
H_{PT}&=&i\kappa \hat{a}^\dagger\hat{a}-i\kappa\hat{b}^\dagger\hat{b}+J(\hat{a}\hat{b}^\dagger+\hat{a}^\dagger\hat{b})
\label{pt}
\end{eqnarray}
for the linear part takes the real eigenvalues when $\kappa<J$; hence the regime of the $\mathcal{PT}$ symmetry under this condition. 
The quantum dynamical properties of this non-Hermitian Hamiltonian of linear system have been recently discussed in \cite{n-h-1, n-h-2}.
Following the same factorization technique in Eq. (\ref{fact}) for the symmetry broken regime with $\kappa>J$, one will find the evolved waveguide modes
\begin{eqnarray}
&&U_{eff}^{-1}(t)\hat{a}U_{eff}(t)= i \frac{\eta_2}{J}e^{-i\zeta_1\chi\int_0^t d\tau e^{2\lambda \tau}\hat{B}_0^\dagger\hat{B}_0(\tau)}\hat{B}_0(t)\nonumber\\
&&U_{eff}^{-1}(t)\hat{b}U_{eff}(t) =  e^{-i\zeta_1\chi\int_0^t d\tau e^{2\lambda \tau} \hat{B}_0^\dagger\hat{B}_0(\tau)}\hat{B}_0(t),
\label{solution-2}
\end{eqnarray}
under the action $U_{eff}(t)=\mathcal{T}e^{-i\int_0^t d\tau (H_{PT}+H_{NL})}$ (see Appendix D), where the operator $\hat{B}_0$ differs from the linear coupler mode operator $\hat{B}$ in (\ref{l-solution}) by neglecting its noise component.
The corresponding expectation value after neglecting the quantum noise drives for the dynamical process in Eq. (\ref{dynamics}) then becomes
\begin{eqnarray}
\langle \hat{b}(t)\rangle=~_a\langle \alpha_0 |e^{-i\zeta_1\chi\int_0^t d\tau e^{4\lambda \tau}\hat{o}_2^\dagger\hat{o}_2}e^{\lambda t}\hat{o}_2 |\alpha_0 \rangle_a. 
\label{noiseless}
\end{eqnarray}
Its difference from the complete quantum expectation value in Eq. (\ref{average}) indicates the nontrivial role of the noises to the system dynamics.

\begin{figure}[t!]
\vspace{0cm}
\centering
\epsfig{file=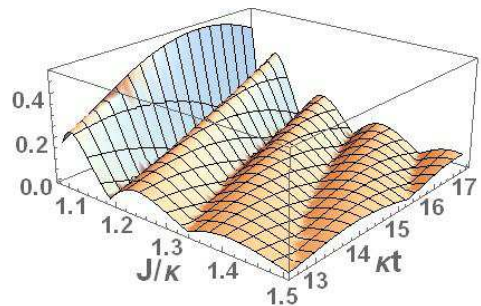,width=0.85\linewidth,clip=} 
{\vspace{-0.2cm}\caption{ (color online) Kerr nonlinearity induced change of a mean quadrature in the $\mathcal{PT}$-symmetric regime. The unmarked vertical axis of the plot is the dimensionless ratio 
$\Delta|\langle \hat{X}_b(\frac{\pi}{2})\rangle|/|\langle \hat{X}^0_b(\frac{\pi}{2})\rangle|$ defined in the text, and its distribution is over the time and in the parameter space. Here we use the quadratic mean of $\langle \hat{X}^0_b(\frac{\pi}{2})\rangle$ over an oscillation period to avoid the singularities from its vanishing values, and the corrections are calculated to the first order of the Kerr coefficient $\chi$ in Eq. (\ref{exact}). 
The system parameters are $\chi=10^{-9}\kappa$ and $\alpha_0=10^3$.  }}
\vspace{0cm}
\end{figure}

\begin{figure}[t!]
\vspace{0cm}
\centering
\epsfig{file=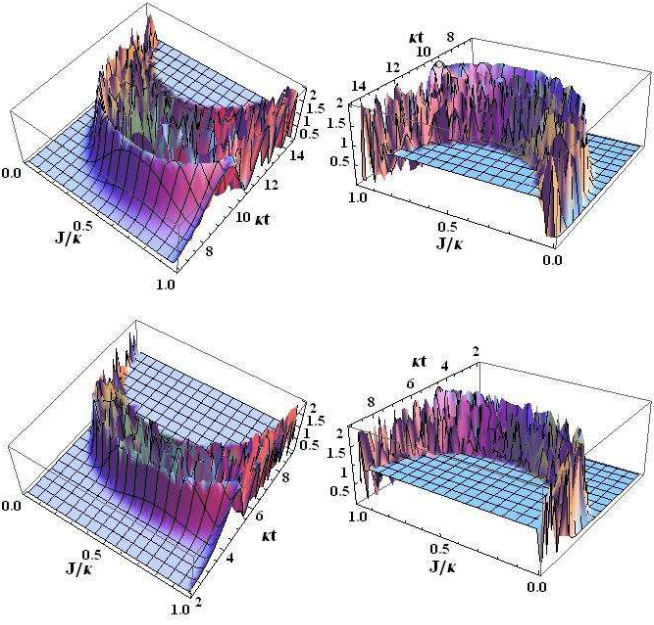,width=1\linewidth,clip=} 
{\vspace{-0.2cm}\caption{(color online) Change of a mean quadrature under the enhanced Kerr nonlinearity in the symmetry broken regime. The 
unmarked vertical axis of the plots is the dimensionless ratio 
$\Delta|\langle \hat{X}_b(\frac{\pi}{2})\rangle|/|\langle \hat{X}^0_b(\frac{\pi}{2})\rangle|$ defined in the text. The upper panels, one as the front view along the time axis and the other as the corresponding back view, are about the 
realistic situation under the quantum noise effects. The lower ones describe the situation under the ``noiseless"  non-Hermitian Hamiltonian (\ref{pt}) together with the Kerr nonlinear action. The platforms of the unit ratio indicate where the mean quadrature disappears due to decoherence. The quantum noises cause the considerable delay of the Kerr nonlinearity enhancement and its consequent decoherence, as compared with the evolution under the non-Hermitian Hamiltonian without noise. Here we take the Kerr coefficient and input coherent state amplitude in Fig. 2.  }}
\vspace{0cm}
\end{figure}

\begin{figure}[t!]
\vspace{0.3cm}
\centering
\epsfig{file=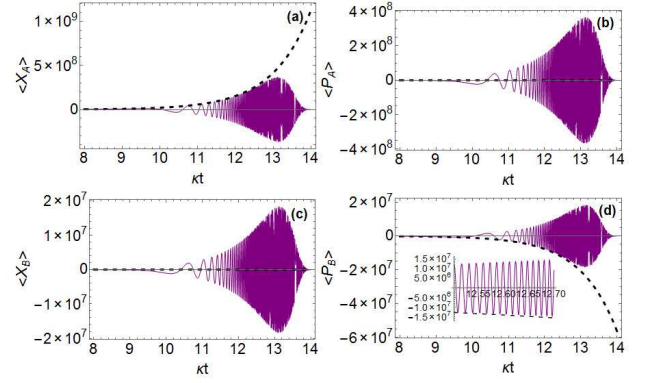,width=1.05\linewidth,clip=} 
{\vspace{-0.3cm}\caption{(color online) Time evolution of the mean quadratures. Here we use the notations $\hat{X}_c(0)=\hat{X}_C$ 
and $\hat{X}_c(\frac{\pi}{2})=\hat{P}_C$, for $C=A$ (the gain channel) or $B$ (the loss channel). The solid curves represent the 
average quadrature values given the Kerr nonlinearity of $\chi=10^{-9}\kappa$, and the dashed curves stand for those of the linear coupler 
with $\chi=0$. The evolutions take place at $J=0.1\kappa$ for the coherent light of $\alpha_0=10^3$. In (d) a refined view of the exponentially accelerating oscillation is shown in the inserted plot. Due to our choice of initially sending the light into the gain channel, the dashed curves coincide with the horizontal axis in (b) and (c).}}
\vspace{0cm}
\end{figure}

\section{Kerr nonlinearity enhancement and quantum noise effects} 

Though it is conceivable that the $\mathcal{PT}$ symmetry breaking will enhance the Kerr effect with the much amplified light intensity, how it works in the system is a main issue we should clarify. For this purpose we illustrate the change ratio $\Delta|\langle \hat{X}_b(\frac{\pi}{2})\rangle|/|\langle \hat{X}^0_b(\frac{\pi}{2})\rangle|$, where $\Delta|\langle \hat{X}_b(\frac{\pi}{2})\rangle|$ is the absolute difference between the mean quadrature $\langle \hat{X}_b(\frac{\pi}{2})\rangle$ under the Kerr effect and 
$\langle \hat{X}^0_b(\frac{\pi}{2})\rangle$ without the Kerr nonlinearity ($\chi=0$). In the $\mathcal{PT}$-symmetric regime of $\kappa<J$, the correction to the average quadrature due to the Kerr nonlinearity can be found by the perturbative expansion according to (\ref{exact}); see Fig. 2. With Eq. (\ref{solution}) we can also obtain the corresponding ratios in the symmetry broken regime. The light field is magnified with the factor $e^{\lambda t}=e^{\sqrt{\kappa^2-J^2}t}$ in the symmetry broken regime, so it would take a longer time  at a larger $J$ to have a considerable change of the mean quadrature by the Kerr nonlinearity. However, the higher coupling rate $J$ enables more light to enter the channel filled with the Kerr medium, making its effect larger. These tendencies combined give rise to the illustrated change ratio distribution in the upper panels of Fig. 3. Surprisingly, there will appear a unit ratio platform (see the upper right panel of Fig. 3), on which the mean quadrature $\langle \hat{X}_b(\frac{\pi}{2})\rangle$ is totally eliminated after a seemingly irregular evolution period. As a contrast, the mean quadrature $\langle \hat{X}(\phi)\rangle$ of the input coherent state can never vanish for arbitrary $\phi$. A decoherence process thus occurs during the time evolution of an input coherent light. One question is whether it is connected with the quantum noise effects boosted in the symmetry broken regime? 

To analyze the $\mathcal{PT}$ symmetry broken regime more thoroughly, we cut across one point on the axis $J/\kappa$ of Fig. 3 to see the evolution of four mean quadratures in Fig. 4. The dashed curves in the figure describe the linear coupler situation \cite{arga}, 
in which the quantum noises do not affect the mean quadrature evolutions at all. There is a pretty symmetry between the mean quadrature evolution in the gain and loss channel, due to the proportionality of the evolved modes in (\ref{solution}). During the beginning period their time evolutions show no difference from those without the Kerr nonlinearity. In addition to deviating these mean quadratures from those of a linear coupler, the gradually enhanced nonlinear action brings an oscillation pattern to their evolutions and, interestingly, the oscillation becomes exponentially fast with time. The seemingly irregular areas of the change ratio distribution in Fig. 3 are where such exponentially accelerating oscillation exists. We can track down its cause in Eq. (\ref{average}). The terms in this equation are proportional to the factor $~_a\langle \alpha_0 |e^{i\zeta_2\chi\int_0^t d\tau e^{2\lambda \tau}\hat{o}_2^\dagger\hat{o}_2} |\alpha_0 \rangle_a$, which is actually the overlap between the input state $|\alpha_0\rangle_a|0\rangle_b$ and the product 
$e^{i\zeta_2\chi\int_0^t d\tau e^{2\lambda \tau}\hat{o}_2^\dagger\hat{o}_2} |\alpha_0 \rangle_a|0\rangle_b=|\beta_1(t)\rangle_a|\beta_2(t)\rangle_b$ of two transformed coherent states. 
Under the broken $\mathcal{PT}$ symmetry giving a real number $\lambda$, the operator-valued phase factor before the input state oscillates at exponentially increasing frequency with time, hence the same behavior of the overlap. 

The overlap $\langle\beta_1(t),\beta_2(t)|\alpha_0,0\rangle$ defined in the above tends to zero with its exponentially accelerating oscillation, leading to the decoherence indicated by the vanishing mean quadratures. This happens because such oscillation from the continually enhanced nonlinear action will instantaneously repeat any close to zero value in the limit of its vanishing oscillation period, and the expectation values $\langle \hat{a}\rangle$ and $\langle \hat{b}\rangle$ will be killed eventually, hence the nonequivalence of the averaged quantum dynamical equations from Eq. (\ref{dynamics}) with the classical dynamical equations 
in Eq. (\ref{c-dynamics}). The mean field approach based on the assumption $\langle \hat{b}^\dagger\hat{b}\rangle=\langle \hat{b}^\dagger\rangle\langle \hat{b}\rangle$ in Eq. (\ref{c-dynamics}) has been used to study the classical aspect (such as light transmission) of the similar nonlinear couplers \cite{n1,n2,n3, n5, n7,n8,n9}. The decoherence phenomenon we have illustrated is irrelevant to the quantum noises, as it also exists in the simplified situation without quantum noise; see the lower panels of Fig. 3. The unit ratio platform in the distribution based on the ``noiseless" expectation value in (\ref{noiseless}) comes into being earlier than those under the noise effects. Due to such decoherence, the light field state of an input coherent state evolving in the symmetry broken regime is very different from a Kerr state \cite{kerr1} generated under a constant Kerr nonlinearity. The coherence of a Kerr state exhibited by its mean quadrature $\langle \hat{X}(\phi)\rangle\neq 0$ with arbitrary $\phi$ revives periodically, but an input coherent light will eventually evolve to a decohered one in our concerned nonlinear coupler.

\begin{figure}[b!]
\vspace{-0.1cm}
\centering
\epsfig{file=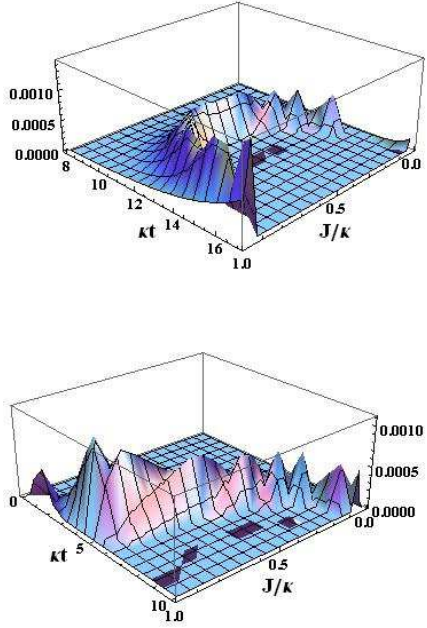,width=0.66\linewidth,clip=} 
{\vspace{-0.2cm}\caption{(color online) Measure of the influence from quantum noise correction on the expectation value of an evolved waveguide mode. The unmarked vertical axis of the plots is the dimensionless quantity $\big|EB_1(t)/EB_0(t)\cdot \langle \beta_1(t),\beta_2(t)|\alpha_0,0\rangle\big|$, which combines the relative intensity of $EB_1(t)$ in Eq. (\ref{average}) and the decoherence effect from the exponentially accelerating oscillation. The upper frame is obtained with the Kerr coefficient $\chi=10^{-9}\kappa$, and lower one is found with $\chi=10^{-5}\kappa$. 
A higher Kerr coefficient does not substantially enhance the effects from the additional term $EB_1(t)$. 
The amplitude of the input coherent state is taken as $\alpha_0=10^3$.  }}
\vspace{-0cm}
\end{figure}

\begin{figure}[h!]
\vspace{0cm}
\centering
\epsfig{file=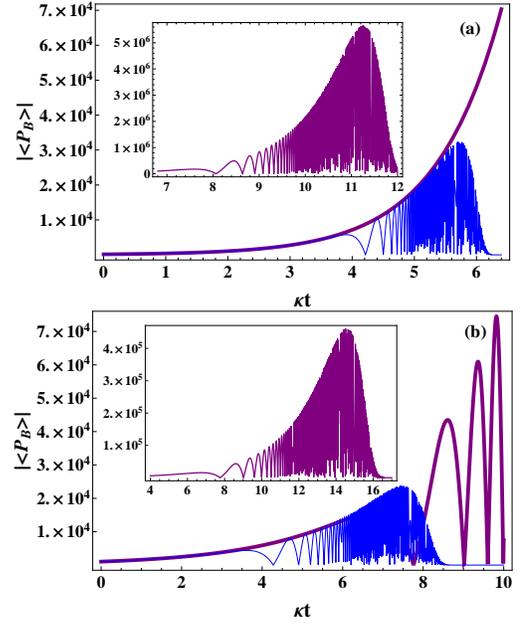,width=0.8\linewidth,clip=} 
{\vspace{-0.3cm}\caption{(color online) Quantum noise induced deviation in a mean quadrature evolution. The thicker (purple) curves stand for the mean quadrature $\hat{X}_b(\frac{\pi}{2})=\hat{P}_B$ under the noise effects, and the thinner (blue) ones are those evolving according to 
the non-Hermitian Hamiltonian in (\ref{pt}) and Kerr nonlinear action. The plots in (a) are found with $J=0.3\kappa$. Those in (b) are obtained with $J=0.9\kappa$. The inserted plot in both (a) and (b) shows a longer time realistic evolution under the noise effects. The Kerr coefficient and input coherent state amplitude are the same as in the previous figures.}}
\vspace{0cm}
\end{figure}

In such open quantum system the noise components will inevitably enter the evolved light field modes as in Eqs. (\ref{exact}) and (\ref{solution}), so some of the system observables can gain the extra contributions from the averages of the involved amplification noise operators over their associated reservoir state. One effect manifested by the evolved photon number operators like this is the spontaneous photon generation, which destroys the nonclassicality of quantum light sent into a linear coupler \cite{arga} but is negligible to our system of strong light fields. For this nonlinear coupler, the system mode operators also gain the extra contributions from the noises, which take the form of the second term $EB_1(t)$ in Eq. (\ref{average}). However, the decoherence from the exponentially accelerating oscillation suppresses their effects when their magnitudes become comparable with the main term $EB_0(t)$; see Fig. 5. The nontrivial way that the quantum noises affect the system dynamics is through their interplay with 
the gradually enhanced Kerr nonlinearity. In each small step of evolution like that with $\delta t\rightarrow 0$ in (\ref{step}), the noise components enter the phase induced by the nonlinearity, while the nonlinear term containing the noise contribution significantly modifies the evolution from that of a linear coupler. As it is manifested by comparing the ``noisy" and ``noiseless" change ratio distributions in Fig. 3, the result is that the quantum noises significantly contribute to the main term $EB_0(t)$ of the expectation values for the evolved waveguide modes, making them totally different from those predicted with the effective Hamiltonian (\ref{pt}) and the Kerr nonlinear term $H_{NL}$. Including the quantum noise effects in the nonlinear coupler, therefore, does not simply add slight modifications to its quantum features. For example, Fig. 6 shows that the quantum noises substantially slow down the change of a mean quadrature by the enhanced Kerr nonlinearity, as compared with that under the system dynamics of the non-Hermitian Hamiltonian excluding the quantum noises. On the other hand, the spontaneous photon generation (independent of the input coherent state amplitude $\alpha_0$) and other corrections to the photon numbers are insignificant in our concerned situation of a strong input coherent beam. 
Accordingly the light intensities proportional to the photon numbers $\langle\hat{a}^\dagger\hat{a}\rangle$ and $\langle\hat{b}^\dagger\hat{b}\rangle$, in which the nonlinearity induced phase is canceled for the main terms, have no considerable difference from those predicted with the effective non-Hermitian Hamiltonian.

\section{Light field entanglement and nonclassicality}
An application of the nonlinear coupler illustrated in Fig. 1 is to generate the macroscopic nonclassical states of strong light fields, since the Kerr effect can be greatly enhanced after breaking the $\mathcal{PT}$ symmetry. One should know the proper regimes for the existence of the nonclassicality such as light field entanglement. Meanwhile it is necessary clarify how the noise effects accompanying light amplification and dissipation will affect the macroscopic nonclassicality. Here we apply the entanglement criterion for CV quantum states in \cite{s-v-2,s-v-3} to find these regimes. In this problem, the criterion based on the negativity of the partially transposed density matrix of this two-mode system can be formulated in terms of the negativity of the third-order moment determinant
\begin{eqnarray}
 \langle \hat{a}^\dagger \hat{a}\rangle\langle \hat{b}^\dagger \hat{b}\rangle D_3&=&\langle \hat{a}^\dagger \hat{a}\rangle\langle \hat{b}^\dagger \hat{b}\rangle+\langle \hat{a}\rangle\langle \hat{b}\rangle
 \langle \hat{a}^\dagger\hat{b}^\dagger\rangle+\langle \hat{a}^\dagger\rangle\langle \hat{b}^\dagger\rangle\langle \hat{a} \hat{b}\rangle \nonumber\\
&-& \langle \hat{a}^\dagger\rangle\langle\hat{a}\rangle\langle \hat{b}^\dagger\hat{b}\rangle-\langle \hat{b}^\dagger\rangle\langle\hat{b}\rangle\langle \hat{a}^\dagger\hat{a}\rangle 
- \langle \hat{a}^\dagger\hat{b}^\dagger\rangle \langle \hat{a}\hat{b}\rangle,\nonumber\\
\label{determinant}
\end{eqnarray}
which is normalized with respect to the product of output photon numbers. Under the full dynamics including the quantum noise effects, the different terms in the above equation are found by averaging the evolved operators over both system and reservoir degrees of freedom. 
Here we present the evolution of the quantity $D_3$ for both situations with and without quantum noises in Fig. 7. 
In both situations the negativity of $D_3$ mainly comes from the difference between the real part of $\langle \hat{a}\rangle\langle \hat{b}\rangle\langle \hat{a}^\dagger\hat{b}^\dagger\rangle$ [due to the second and third term in (\ref{determinant})] and 
the absolute value $\langle \hat{a}^\dagger\rangle\langle\hat{a}\rangle\langle \hat{b}^\dagger\hat{b}\rangle$ or $\langle \hat{b}^\dagger\rangle\langle\hat{b}\rangle\langle \hat{a}^\dagger\hat{a}\rangle $, which appears in the progress of enhancing the Kerr nonlinearity. Inside the proper regime the realized macroscopic entangled states can have more than $10^{14}$ photons in 
the example of Fig. 4. Thanks to the proportionality between the evolved modes in (\ref{solution}), the negativity of $D_3$ happens to be the necessary and sufficient condition for the general nonclassicality of the evolved light field states, when their $P$ functions fail to be classical probability distributions \cite{s-v-1}. The macroscopic nonclassical states can be 
used to generate other types of entanglement by coupling them with different photonic states via a beam-splitter \cite{bs1,bs2}.
Due to the decoherence mechanism we have discussed before, all terms except for the photon numbers in (\ref{determinant}) will be killed with time, leaving the increase of the photon numbers $ \langle\hat{a}^\dagger\hat{a}\rangle $ and  $\langle\hat{b}^\dagger\hat{b}\rangle $ to be the single process in the end.    

\begin{figure}[t!]
\vspace{0cm}
\centering
\epsfig{file=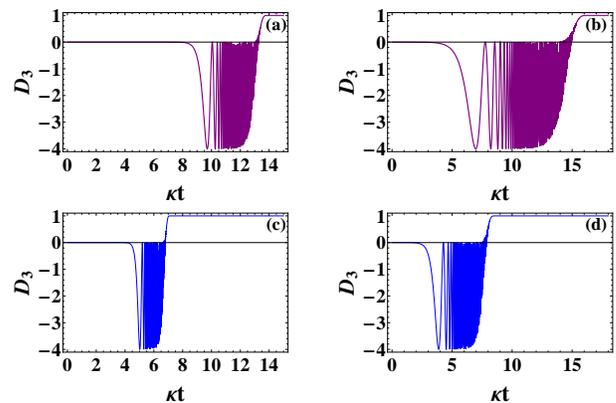,width=1\linewidth,clip=} 
{\vspace{-0.3cm}\caption{(color online) Emergence of the entanglement and nonclassical states. The negative values of $D_3$ indicate the entanglement of the output fields and their general nonclassicality. (a) and (b) show the quantity evolution under the full dynamics including the quantum noise effects. The system parameters are taken as $J=0.1\kappa$ in (a) and $J=0.9\kappa$ in (b), respectively. (c) illustrates the corresponding evolution to (a), but is due to the non-Hermitian Hamiltonian (\ref{pt}) plus the nonlinear Hamiltonian 
$H_{NL}$ without quantum noise. (d) is the corresponding situation for (b) without the quantum noise effects as well. The plot in (b) shows that the nonclassical regime can be realized with an amplification rate $e^{\lambda t}\sim 10$. All other parameters are the same as in Fig. 4.}}
\vspace{0cm}
\end{figure}

As seen from Fig. 7, the regimes for realizing the nonclassical states are more extended under the effects of the quantum noises. This is very different from the phenomena of destroying the nonclassicality by quantum noises in other systems (see, e.g. \cite{esd}).
The nonclassicality occurs with the enhanced Kerr nonlinearity, which will finally lead to a decoherence. Like the other nonlinearity determined features, the meaningful regimes for generating the nonclassical states are also characterized by the expectation value of the nonlinearity induced phase factor; $~_a\langle \alpha_0 |e^{i\zeta_2\chi\int_0^t d\tau e^{2\lambda \tau}\hat{o}_2^\dagger\hat{o}_2} |\alpha_0 \rangle_a$ for the realistic situation under the noise effects, and $~_a\langle \alpha_0 |e^{-i\zeta_1\chi\int_0^t d\tau e^{4\lambda \tau}\hat{o}_2^\dagger\hat{o}_2} |\alpha_0 \rangle_a$ for the simplified situation determined by the non-Hermitian effective Hamiltonian. The nonlinearity induced phase in the former situation can not exist without the quantum noises, since it is proportional to the coefficient $\zeta_2$ originating from the noise action. The nonclassical states of the output fields are therefore effectively created by the joint action of the quantum noises and enhanced nonlinearity. Two effects, the enhanced nonlinearity and the accompanying decoherence, indicated by these expectation values of the phase factors are crucial to the generation of the macroscopic nonclassicality. The nonclassical states exist in where the Kerr nonlinearity is sufficiently large but the decoherence has not been significant. By slowing down the pace toward the decoherence consequent to the Kerr nonlinearity enahncement, the quantum noises help to achieve a wider time window for the existence of the nonclassical states. This phenomenon is similar to the synergic effects of noise and nonlinearity in classical systems. A typically analogous example is stochastic resonance \cite{sr}, the phenomenon that input classical noises help to improve output signal to noise ratios. Such effect of the quantum noises can be controlled by the parameter $J/\kappa$. With a parameter $J/\kappa$ chosen as that in Fig. 7(b), it is possible to reach the regime of the nonclassical states with a moderate amplification rate $e^{\lambda t}$. 

\section{discussion}

We have found the solution to the dynamical problem about a coherent light evolving in the nonlinear coupler in Fig. 1, focusing on the effects from the enhanced Kerr nonlinearity, as well as the inherent quantum noises. By appearance the enhancement of Kerr nonlinearity due to much strengthened light intensity in the symmetry broken regime is a rather straightforward result. When it comes to its quantum properties, however, the system exhibits rich and unexpected phenomena related to this Kerr nonlinearity enhancement. The coexistence of field coupling plus amplification and dissipation, nonlinearity, as well as quantum noises, makes this simple system a unique platform for studying complicated quantum dynamics. These factors build up the highly symmetric and unusual evolution of the mean quadratures of the waveguide modes. Instead of originating from the quantum noises, the decoherence of an input coherent light is caused by the enhanced Kerr nonlinearity itself. The regimes for the macroscopic entanglement of light fields are found to exist in where the Kerr nonlinearity is enhanced but has not reached the degree of giving rise to the decoherence. Such decoherence determines the final 
state of an input coherent light.

In this work we adopt a single-mode description for the light fields, so that the form of dynamical equations Eq. (\ref{dynamics}) should correspond to the mean field format Eq. (\ref{c-dynamics}) used among the literature. The real light fields in waveguides are multi-mode. The multi-mode nature of light fields gives rise to important effects on the evolved light field quantum states out of Kerr nonlinearity \cite{mc-1,mc-2,mc-3,mc-4,mc-5,mc-6,mc-7, mc-8}. However, the effects we consider here are measured with the relevant expectation values, and are not directly connected with the evolved quantum states. We work in the Heisenberg picture to find the expectation values from the input light field state. The qualitative picture of the interplay between the enhanced nonlinearity and quantum noises will not be changed with a multi-mode description for the light fields; see Appendix E. The essential features in the process involving these physical elements can be well captured by the single-mode dynamical equations, so the evolution of single modes of waveguides is sufficient for illustrating the concerned phenomena.  

We also assume a constant amplification rate $\kappa$ in the discussions. When the light intensity becomes sufficiently large, especially in the symmetry broken regime, the amplification rate will be lowered to $\kappa/(1+I/I_s)$, where $I$ is proportional to $\langle\hat{a}^\dagger\hat{a}\rangle$ and $I_s$ is the saturation intensity. The influence of such amplification saturation on  coupled gain-loss cavities under external driving field has been experimentally demonstrated \cite{ex4}. In our system it will modify 
the dynamical equation of the gain channel mode in Eqs. (\ref{dynamics}) and (\ref{c-dynamics}). $\mathcal{PT}$ symmetric regime 
will no longer exist under its effect, and the reduced amplification compared with the unchanged dissipation makes the system of coupled waveguide modes tend to a steady state of vanishing observables. With the constantly equal amplification and dissipation rate, the linear part of the system exhibits a clear-cut ``phase transition'' between oscillating and exponentially growing waveguide modes across the threshold $\kappa=J$, so it is more meaningful and transparent to illustrate how the action of an added tiny nonlinearity should change in this situation of the balanced gain and loss. Given a sufficiently large saturation intensity $I_s$, the obviously different system observables from our illustrations are expected only after the evolution of light fields in the symmetry broken regime for a considerable period of time.     

The most important finding of the study is about the effects of the quantum noises in the $\mathcal{PT}$ symmetry broken regime, where the coupler is a typical quantum system with non-perturbative nonlinearity. Quantum systems with both non-perturbative nonlinearity and noises were generally difficult to deal with by the standard methods. The approach presented here makes it feasible to find the physical observables of the related intricate systems with both nonlinearity and quantum noises, when there is also no steady-state solution for the dynamical processes. Instead of perturbatively modifying the results predicted by the non-Hermitian effective Hamiltonian, the quantum noises substantially influence the dynamics of the nonlinear coupler. Another interesting phenomenon involving the quantum noises is their creation of the nonclassical output states when they act together with the enhanced nonlinearity. The regime for the nonclassical states is also larger than the prediction with the ``noiseless" non-Hermitian Hamiltonian, as the quantum noises slow down the decoherence growing with the Kerr nonlinearity enhancement. This is analogous to various synergic phenomena involving classical noises in nonlinear systems. In stark contrast to the physics of classical noises that has been well explored in classical nonlinear systems, the research on quantum noises in the systems with non-perturbative nonlinearity is only at the beginning stage. The current work may stimulate the development in this direction.

\begin{acknowledgments}

M.X. acknowledges funding support in part from NBRPC (Grant No. 2012CB921804) and NSFC (No. 61435007 and No. 11321063). S.-B. Yan is supported by NSFC (No. 91123036, No. 61178058, and No. 61275166).

\end{acknowledgments}

\vspace{0cm}

\appendix

\subsection* {Appendix A: Properties of the stochastic Hamiltonian}

\renewcommand{\theequation}{A-\arabic{equation}}
\renewcommand{\thefigure}{A-\arabic{figure}}
\setcounter{equation}{0}
\setcounter{figure}{0}

\renewcommand{\theequation}{A-\arabic{equation}}
\setcounter{equation}{0}

First, we gives a brief explanation of the construction of the stochastic Hamiltonian in Eq. (\ref{H}) of the main text. 
The reservoir for the dissipation of light is modeled as an ensemble of oscillators with the positive energy, and that for the light amplification is, however, an ensemble of oscillators with the negative energy \cite{book}. The reservoirs are in the vacuum states at the assumed zero temperature. With respect to the self oscillation Hamiltonian $H_0=\omega_a \hat{a}^\dagger \hat{a}+\omega_b \hat{b}^\dagger \hat{b}$ of the waveguide modes and that of the reservoirs $H_R=-\int d\omega \omega \hat{\xi}^\dagger_a(\omega)\hat{\xi}_a(\omega)+
\int d\omega \omega \hat{\xi}^\dagger_b(\omega)\hat{\xi}_b(\omega)$, the general coupling Hamiltonian between the system 
and reservoirs takes the form
\begin{eqnarray}
H_{int}&=&i\int_{-\infty}^{\infty}d\omega \gamma_1(\omega)\big(\hat{\xi}_a(\omega)e^{i\omega t}+\hat{\xi}_a^\dagger(\omega)e^{-i\omega t}\big)\nonumber\\
&\times & \big(\hat{a}^\dagger e^{i\omega_a t}-\hat{a}e^{-i\omega_a t}\big)\nonumber\\
&+&i\int_{-\infty}^{\infty}d\omega \gamma_2(\omega)\big(\hat{\xi}_b(\omega)e^{-i\omega t}+\hat{\xi}_b^\dagger(\omega)
e^{i\omega t}\big)\nonumber\\
&\times & \big(\hat{b}^\dagger e^{i\omega_b t}-\hat{b}e^{-i\omega_b t}\big)
\end{eqnarray}
in the interaction picture. Taking the substitution $\gamma_1(\omega)$, $\gamma_2(\omega) \rightarrow \sqrt{2\kappa/(2\pi)}$ of the system-reservoir couplings in our concerned situation of the balanced gain and loss, and applying the rotation wave approximation (RWA) that neglects the fast oscillating terms in the above equation, the coupling Hamiltonian can be rewritten as
\begin{eqnarray}
H_{int}&=&i\sqrt{2\kappa}\{\hat{a}^\dagger \hat{\xi}^\dagger_a(t)-\hat{a} \hat{\xi}_a(t)\}\nonumber\\
&+&i\sqrt{2\kappa}\{\hat{b}^\dagger \hat{\xi}_b(t)-\hat{b} \hat{\xi}^\dagger_b(t)\},
\end{eqnarray}
where 
$$\hat{\xi}_a(t)=\frac{1}{\sqrt{2\pi}}\int_{-\infty}^{\infty} d\omega \hat{\xi}_a(\omega) e^{i(\omega-\omega_a)t}, $$ 
$$\hat{\xi}_b(t)=\frac{1}{\sqrt{2\pi}}\int_{-\infty}^{\infty} d\omega \hat{\xi}_b(\omega) e^{-i(\omega-\omega_b)t}. $$ 

Corresponding to the correlation relations 
$$\langle \hat{\xi}^\dagger_c(t)\hat{\xi}_c(t')\rangle=0,~~~ \langle \hat{\xi}_c(t)\hat{\xi}_c^\dagger(t')\rangle=\delta(t-t')$$ 
for the noise operators with $c=a$ and $b$, there are the Ito's rules
\begin{eqnarray}
&&\big(d\hat{W}_i(t)\big)^2=\big(d\hat{W}^\dagger_i(t)\big)^2=0,\nonumber\\
&&d\hat{W}_i(t)d\hat{W}^\dagger_j(t)=\delta_{ij} dt,\nonumber\\
&& d\hat{W}^\dagger_i(t)d\hat{W}_j(t)=d\hat{W}^\dagger_i(t)d\hat{W}_i(t)=0
\label{Ito}
\end{eqnarray}
for the defined stochastic operator $\hat{W}_i(t)=\int_0^t d\tau \hat{\xi}_i(\tau)$. Using the notation $H_S=J(\hat{a}\hat{b}^\dagger+\hat{a}^\dagger\hat{b})+1/2\chi(\hat{b}^\dagger)^2\hat{b}^2$ for the system Hamiltonian of the nonlinear coupler, one will have the increment of the waveguide mode operators under the joint action $U(t+dt,t)=e^{-i\{H_L(t)+H_{NL}\}dt}$ as
\begin{eqnarray}
d\hat{c}(t)&=&U^\dagger(t+dt,t)\hat{c}(t)U(t+dt,t)-\hat{c}(t)\nonumber\\
&=& i[H_S,\hat{c}]dt-\sqrt{2\kappa}[\hat{a}^\dagger d\hat{W}^\dagger_a(t)-\hat{a} d\hat{W}_a(t),\hat{c}]\nonumber\\
&-&\sqrt{2\kappa}[\hat{b}^\dagger d\hat{W}_b(t)-\hat{b} d\hat{W}^\dagger_b(t),\hat{c}]\nonumber\\
&+& \kappa (2\hat{a}\hat{c}\hat{a}^\dagger-\hat{c}\hat{a}\hat{a}^\dagger-\hat{a}\hat{a}^\dagger \hat{c})dt\nonumber\\
&+& \kappa (2\hat{b}^\dagger\hat{c}\hat{b}-\hat{c}\hat{b}^\dagger\hat{b}-\hat{b}^\dagger\hat{b} \hat{c})dt,
\end{eqnarray}
where the small element $U(t+dt,t)$ of the joint action defined in Eq. (\ref{fact}) of the main text is expanded to the second order, and the Ito's rules in (\ref{Ito}) are applied to the involved stochastic operators. It is straightforward to obtain the system dynamical equations in Eq. (\ref{dynamics}) of the main text, after substituting the waveguide mode operator $\hat{c}=\hat{a}$ or $\hat{b}$ into the above equation. 

In the Schr\"{o}dinger picture, the increment from the action $U(t+dt,t)$ on a joint state $\hat{\rho}(t)$ of the system plus reservoirs is as follows:
\begin{eqnarray}
d\hat{\rho}(t)&=&U(t+dt,t)\hat{\rho}(t)U^\dagger(t+dt,t)-\hat{\rho}(t)\nonumber\\
&=&-i\big\{(H_{C}+H_{NL})\hat{\rho}(t)-\hat{\rho}(t)(H^{\dagger}_{C}+H_{NL})\big\}dt\nonumber\\
&+&2\kappa \big \{d\hat{W}^\dagger_a(t)\hat{a}^\dagger\hat{\rho}(t)\hat{a} d\hat{W}_a(t)
+d\hat{W}^\dagger_b(t)\hat{b}\hat{\rho}(t) \hat{b}^\dagger d\hat{W}_b(t)\big\}\nonumber\\
&+&\sqrt{2\kappa}\big\{d \hat{W}^\dagger_a \hat{a}^\dagger \hat{\rho}(t)+\hat{\rho}(t) d\hat{W}_a\hat{a}\big\}\nonumber\\
&+& \sqrt{2\kappa}\big\{d \hat{W}^\dagger_b \hat{b}\hat{\rho}(t)+\hat{\rho}(t) d\hat{W}_b\hat{b}^\dagger\big\},
\label{master}
\end{eqnarray}
where the effective Hamiltonian for the linear part is 
 \begin{eqnarray}
H_C=-i\kappa \hat{a}\hat{a}^\dagger-i\kappa\hat{b}^\dagger\hat{b}+J(\hat{a}\hat{b}^\dagger+\hat{a}^\dagger\hat{b}).
\end{eqnarray}
Tracing out the reservoir degrees of freedom in this equation, while considering the Ito's rules in (\ref{Ito}), one will obtain the quantum master equation [Eq. (\ref{m-e}) in the main text] about the reduced density matrix $\rho(t)=\mbox{Tr}_R\hat{\rho}(t)$ of the system. 
 
\subsection* {Appendix B: Expansion of the nonlinear action}

\renewcommand{\theequation}{B-\arabic{equation}}
\renewcommand{\thefigure}{B-\arabic{figure}}
\setcounter{equation}{0}
\setcounter{figure}{0}

\renewcommand{\theequation}{B-\arabic{equation}}
\setcounter{equation}{0}

The joint evolution operator $U(t)$ as the time-ordered exponential $\mathcal{T}e^{-i\int_0^t d\tau (H_L+H_{NL})(\tau)}$ for the 
nonlinear coupler satisfies the following differential equation
\begin{eqnarray}
\frac{d U(t)}{dt}=-i\big(H_L(t)+H_{NL}\big)U(t),
\end{eqnarray}
with the initial condition $U(0)=I$, the unit operator. 
On the other hand, the process $U_L(t)$ solely under the first linear Hamiltonian $H_L$ is the solution of the differential equation
\begin{eqnarray}
\frac{d U_L(t)}{dt}
&=&-iH_L(t)U_L(t)
\end{eqnarray}
with the same initial condition. Two variations of these equations are the quantum stochastic differential equations (QSDE) 
in Stratonovich and Ito form \cite{book}. We consider the following differential
\begin{eqnarray}
&&\frac{d}{dt}\big\{U^\dagger_L(t)U(t)\big\}\nonumber\\
&=&iU^\dagger_L(t)H_L(t)U(t)+U^\dagger_L(t)(-iH_L(t)-iH_{NL})U(t)\nonumber\\
&=&-iU^\dagger_L(t)H_{NL}U(t).
\label{tt}
\end{eqnarray}
The integral of the above equation from $0$ to $t$ gives
\begin{eqnarray}
U(t)=U_L(t)\big (I-i\int_0^t ds_1 U^\dagger_L(s_1)H_{NL}U(s_1)\big).
\end{eqnarray}
Substituting the same expression for the operator $U(s_1)$ into the above leads to the second order formula
\begin{eqnarray}
&&U(t)=U_L(t)\big (I+(-i)\int_0^t ds_1 U^\dagger_L(s_1)H_{NL}U_L(s_1)\nonumber\\
&-&\int_0^t ds_1 U^\dagger_L(s_1)H_{NL}U_L(s_1)\int_0^{s_1} ds_2 U^\dagger_L(s_2)H_{NL}U(s_2)\big).\nonumber\\
\end{eqnarray}
Iteratively applying the above procedure in the above equation, one will obtain the expansion in Eq. (\ref{fact}) of the main text.
In the symmetry broken regime where the transformed Hamiltonian $H'_{NL}(t)=1/2\chi(\hat{B}^\dagger(t))^2(\hat{B}(t))^2$ becomes non-perturbative, each term in the expansion should be well taken into account for the dynamical process. 

\vspace{0.3cm}
\subsection* {Appendix C: Non-perturbative solution in the symmetry broken regime}

\renewcommand{\theequation}{C-\arabic{equation}}
\renewcommand{\thefigure}{C-\arabic{figure}}
\setcounter{equation}{0}
\setcounter{figure}{0}

Eq. (\ref{fact}) in the main text enables one to find the evolved mode operators. The first step is with the linear action $U_L(t)$ to get
\begin{widetext}
\begin{eqnarray}
&&U^\dagger_L(t)\hat{a}U_L(t)=\hat{A}(t)=-ie^{-\lambda t}\frac{\eta_1}{J}(i\frac{J}{\eta_1+\eta_2}\hat{a}+\frac{\eta_2}{\eta_1+\eta_2}\hat{b})+i
e^{\lambda t}\frac{\eta_2}{J}(-i\frac{J}{\eta_1+\eta_2}\hat{a}+\frac{\eta_1}{\eta_1+\eta_2}\hat{b})\nonumber\\
&+&\sqrt{2\kappa}\int_0^t d\tau \big (-ie^{-\lambda (t-\tau)}\frac{\eta_1}{J}\big(i\frac{J}{\eta_1+\eta_2}\hat{\xi}_a^\dagger(\tau)+\frac{\eta_2}{\eta_1+\eta_2}\hat{\xi}_b(\tau)\big)+i
e^{\lambda (t-\tau)}\frac{\eta_2}{J}\big(-i\frac{J}{\eta_1+\eta_2}\hat{\xi}_a^\dagger(\tau)+\frac{\eta_1}{\eta_1+\eta_2}\hat{\xi}_b(\tau)\big),\nonumber\\
&&U^\dagger_L(t)\hat{b}U_L(t)=\hat{B}(t)=e^{-\lambda t}(i\frac{J}{\eta_1+\eta_2}\hat{a}+\frac{\eta_2}{\eta_1+\eta_2}\hat{b})+e^{\lambda t}\underbrace{(-i\frac{J}{\eta_1+\eta_2}\hat{a}+\frac{\eta_1}{\eta_1+\eta_2}\hat{b})}_{\hat{o}_2}\nonumber\\
&+&\sqrt{2\kappa}\int_0^t d\tau \big(e^{-\lambda (t-\tau)}\big(i\frac{J}{\eta_1+\eta_2}\hat{\xi}_a^\dagger(\tau)+\frac{\eta_2}{\eta_1+\eta_2}\hat{\xi}_b(\tau)\big)+\underbrace{\sqrt{2\kappa}\int_0^t d\tau e^{\lambda (t-\tau)}\big(-i\frac{J}{\eta_1+\eta_2}\hat{\xi}^\dagger_a(\tau)+\frac{\eta_1}{\eta_1+\eta_2}\hat{\xi}_b(\tau)\big)}_{\hat{n}(t)},\nonumber\\
\label{linear}
\end{eqnarray}
\end{widetext}
where $\eta_{1(2)}=\mp\kappa+ \sqrt{\kappa^2-J^2}$ and $\lambda= \sqrt{\kappa^2-J^2}$. In dealing with the noise operators we have applied
the Ito's rules in (\ref{Ito}).

The next step is the nonlinear action $U_{NL}(t)=\mathcal{T}e^{-i\int_0^t d\tau U^\dagger_L(\tau)H_{NL}U_L(\tau)}$. Before we apply the action to the linear coupler modes in (\ref{linear}), we give a useful commutator between the transformed number operators:
\begin{eqnarray}
&&[\hat{B}^\dagger\hat{B}(t),\hat{B}^\dagger\hat{B}(t')]=[e^{2\lambda t}\hat{o}^\dagger_2\hat{o}_2,e^{2\lambda t'}\hat{o}^\dagger_2\hat{o}_2]\nonumber\\
&+& e^{\lambda t}\hat{o}^\dagger_2\underbrace{[\hat{n}(t),\hat{n}^\dagger(t')]}_{\epsilon_1(t,t')}e^{\lambda t'}\hat{o}_2
+e^{\lambda t'}\hat{o}^\dagger_2\underbrace{[\hat{n}^\dagger(t),\hat{n}(t')]}_{-\epsilon_1(t,t')}e^{\lambda t}\hat{o}_2\nonumber\\
&+&\hat{n}^\dagger(t)\underbrace{[e^{\lambda t}\hat{o}_2,e^{\lambda t'}\hat{o}_2^\dagger]}_{\epsilon_2(t,t')}\hat{n}(t')
+\hat{n}^\dagger(t')\underbrace{[e^{\lambda t}\hat{o}_2^\dagger,e^{\lambda t'}\hat{o}_2]}_{-\epsilon_2(t,t')} \hat{n}(t)\nonumber\\
&+&\hat{n}^\dagger(t)[\hat{n}(t),\hat{n}^\dagger(t')]\hat{n}(t')
+\hat{n}^\dagger(t')[\hat{n}^\dagger(t),\hat{n}(t')] \hat{n}(t)\nonumber\\
&\equiv & \hat{N}(t,t'),
\label{n-c}
\end{eqnarray}
where the terms with the decay factor in (\ref{linear}) are neglected.
The above commutator is a pure noise operator $ \hat{N}(t,t')$ because the part containing the system operators is canceled. The expectation value of the operator $\hat{B}^\dagger\hat{B}(t)\big[\big(\hat{B}^\dagger(\tau)\big)^2\big(\hat{B}(\tau)\big)^2,\hat{B}(t)\big]$ is therefore much larger than that of 
$\big[\big(\hat{B}^\dagger(\tau)\big)^2\big(\hat{B}(\tau)\big)^2,\hat{B}^\dagger\hat{B}(t)\big]\hat{B}(t)$, given a strong input coherent light 
with $\alpha_0\gg 1$. The former is in the order of $\alpha_0^5$ but the latter is proportional to $\alpha_0^3$. This relation will be used below.

\vspace{0cm}
With the infinite product form 
\begin{eqnarray}
&&U_{NL}(t)=\mathcal{T}e^{-i\int_0^t d\tau H'_{NL}(\tau)}\nonumber\\
&=& \lim_{n\rightarrow \infty} e^{-iH'_{NL}(t_{n})\delta t}\cdots  e^{-iH'_{NL}(t_2)\delta t}e^{-iH'_{NL}(t_1)\delta t}\nonumber\\
\label{infinite-0}
\end{eqnarray} 
of the time ordered exponential, where $H'_{NL}(t)=1/2\chi(\hat{B}^\dagger(\tau))^2(\hat{B}(\tau))^2$ and the time range $\tau\in [0,t]$ is divided into $n$ pieces with $n\rightarrow \infty$, we perform the transformation by the nonlinear action as follows:  
\begin{widetext}
\begin{eqnarray}
\hat{a}(t)&=& U^\dagger_{NL}(t)\hat{A}(t)U_{NL}(t)=e^{iH'_{NL}(t_{1})\delta t}\cdots  e^{iH'_{NL}(t_{n-1})\delta t}e^{iH'_{NL}(t_n)\delta t}\hat{A}(t)e^{-iH'_{NL}(t_n)\delta t} e^{-iH'_{NL}(t_{n-1})\delta t}\cdots e^{-iH'_{NL}(t_{1})\delta t}\nonumber\\
&=&   e^{iH'_{NL}(t_{1})\delta t}\cdots  e^{iH'_{NL}(t_{n-1})\delta t} \big(\hat{A}(t) -i\chi c_{ab}(t,t_n)\hat{B}^\dagger(t_n)\hat{B}(t_n)\hat{B}(t_n)\delta t\big)e^{-iH'_{NL}(t_{n-1})\delta t}\cdots e^{-iH'_{NL}(t_{1})\delta t}\nonumber\\
&=& \hat{A}(t)-i\chi  c_{ab}(t,t_n)U^\dagger_{NL}(t_n,t_1)\hat{B}^\dagger\hat{B}\hat{B}(t_n)U_{NL}(t_n,t_1)\delta t-\cdots -i\chi  c_{ab}(t,t_{2})U^\dagger_{NL}(t_2,t_{1})\hat{B}^\dagger\hat{B}\hat{B}(t_{2})U_{NL}(t_2,t_{1})\delta t\nonumber\\
&-& i\chi  c_{ab}(t,t_{1})\hat{B}^\dagger\hat{B}\hat{B}(t_{1})\delta t\nonumber\\
&=& \hat{A}(t)-i\chi \int_0^t d\tau_1 c_{ab}(t,\tau_1) U^\dagger_{NL}(\tau_1)\hat{B}^\dagger(\tau_1)\hat{B}(\tau_1)\hat{B}(\tau_1)
U_{NL}(\tau_1).
\label{m1}
\end{eqnarray}
\end{widetext} 
The infinite sum ($n\rightarrow \infty$) after the second last equality sign of the above equation is equivalent to the integral 
on the last line, and each term in the summation is modified by the 
preceding action $U_{NL}(t_i,t_1)=U_{NL}(t_i)$ ($t_1=0$) before the moment $t_i$.
In the commutator 
\begin{eqnarray}
&&c_{ab}(t,t')=[\hat{A}(t),\hat{B}^\dagger(t')]\nonumber\\
&=& i\frac{\eta_2}{J}(\zeta_1+\zeta_2)e^{\lambda t+\lambda t'}
-i\frac{\eta_2}{J}\zeta_2e^{\lambda |t-t'|}
\label{terms}
\end{eqnarray}
defined in the above equation, the factor $\zeta_1=\frac{\eta_1^2+J^2}{(\eta_1+\eta_2)^2}$ arises from the exponentially increasing term $e^{\lambda t}\hat{o}_2$ in (\ref{linear}), while the noise 
operator $\hat{n}(t)$ defined in (\ref{linear}) gives two terms $i\frac{\eta_2}{J}\zeta_2e^{\lambda t+\lambda t'}$ and $-i\frac{\eta_2}{J}\zeta_2e^{\lambda |t-t'|}$ with $\zeta_2=\frac{\kappa}{\lambda}\frac{\eta_1^2-J^2}{(\eta_1+\eta_2)^2}$. 
The first one cancels the contribution from the system operators with the relation $\zeta_1+\zeta_2=0$. This relation guarantees the equal-time commutation relation $[\hat{A}(t),\hat{A}^\dagger(t)]=[\hat{B}(t),\hat{B}^\dagger(t)]=1$ for the exact linear coupler modes in (\ref{linear}).

The integral in the second term on the last line of (\ref{m1}) can be approximated by
$$\int_0^t d\tau_1 c_{ab}(t,\tau_1) \hat{B}^\dagger(\tau_1)\hat{B}(\tau_1)U^\dagger_{NL}(\tau_1)\hat{B}(\tau_1)U_{NL}(\tau_1),$$
i.e. the operator $\hat{B}^\dagger\hat{B}(\tau_1)$ and the action $U_{NL}(\tau_1)$ are approximately commutative.
As we have discussed following Eq. (\ref{n-c}), for a strong input coherent light, the result from the further nonlinear action on the transformed number operator $\hat{B}^\dagger\hat{B}(\tau_1)$ contributes to a negligible expectation value on the same order of the Kerr coefficient 
$\chi$, as compared with the integral in the above. Then the evolved mode will be expanded to all orders of the Kerr coefficient $\chi$ as follows:
\begin{widetext}
\begin{eqnarray}
\hat{a}(t)
&=& \hat{A}(t)-i\chi \int_0^t d\tau_1 c_{ab}(t,\tau_1) \hat{B}^\dagger(\tau_1)\hat{B}(\tau_1)U^\dagger_{NL}(\tau_1,0)\hat{B}(\tau_1)U_{NL}(\tau_1,0)\nonumber\\
&=& \hat{A}(t)-i\chi \int^t_0 d\tau_1 c_{ab}(t,\tau_1) \hat{B}^\dagger\hat{B}\hat{B}(\tau_1)+(-i\chi)^2\int_0^t d\tau_1 c_{ab}(t,\tau_1) \hat{B}^\dagger\hat{B}(\tau_1)\int_{0}^{\tau_1}d\tau_2 c_{bb}(\tau_1,\tau_2) \hat{B}^\dagger\hat{B}\hat{B}(\tau_2)+\cdots\nonumber\\
&+& (-i\chi)^n \int_0^t d\tau_1 c_{ab}(t,\tau_1) \hat{B}^\dagger\hat{B}(\tau_1)\int_{0}^{\tau_1}d\tau_2 c_{bb}(\tau_1,\tau_2) \hat{B}^\dagger \hat{B}(\tau_2)\cdots \int_{0}^{\tau_{n-1}}d\tau_n c_{bb}(\tau_{n-1},\tau_n) \hat{B}^\dagger\hat{B}\hat{B}(\tau_n)+\cdots\nonumber\\
\label{m2}
\end{eqnarray}
\end{widetext} 
After neglecting the decay terms in (\ref{linear}), the commutator $c_{bb}(t,t')=[\hat{B}(t),\hat{B}^\dagger(t')]$ 
has the relation $c_{ab}(t,t')=i\frac{\eta_2}{J}c_{bb}(t,t')$, and
\begin{eqnarray}
[\hat{B}(\tau_{i}),\hat{B}^\dagger(\tau_{i+1})]
= -\zeta_2 e^{\lambda |\tau_i-\tau_{i+1}|}
=-\zeta_2 e^{\lambda (\tau_{i}-\tau_{i+1})}~~~
\end{eqnarray}
because $\tau_{i}>\tau_{i+1}$ ($1\leq i<\infty$) for each term $c_{bb}(\tau_i,\tau_{i+1})$ of (\ref{m2}).
Plugging these commutators $c_{ab}$ and $c_{bb}$ into (\ref{m2}) reduces the evolved mode to
\begin{eqnarray}
\hat{a}(t)
&=&i\frac{\eta_2}{J} \hat{B}(t)+i\frac{\eta_2}{J} \left ( \mathcal{T}e^{i\zeta_2\chi\int_0^t d\tau \hat{B}^\dagger\hat{B}(\tau)}-I  \right)\hat{B}(t)\nonumber\\
&+&\sum_{n=1}^\infty(i\zeta_2\chi)^n\int_0^t d\tau_1 \hat{B}^\dagger\hat{B}(\tau_1)\int_0^{\tau_{1}}d\tau_2 \hat{B}^\dagger\hat{B}(t_2)\cdots\nonumber\\
&\times & \int_0^{\tau_{n-1}}d\tau_n \hat{B}^\dagger\hat{B}(\tau_n)\hat{\nu}(t,\tau_{n}),
\label{mode-noise}
\end{eqnarray}
where $\hat{\nu}(t,\tau_n)=\sqrt{2\kappa}\int_t^{\tau_n}dt' e^{\lambda(t-t')}\hat{n}_2(t')$. In deriving the above result the exponential factor $e^{\lambda t}$ in the first commutator $c_{ab}(t,\tau_1)$ of each term (except for the first one) on the right side of (\ref{m2}) is moved to the most right operator $\hat{B}(\tau_n)$. With the noise operator $\hat{\nu}(t,\tau_n)$ in each term to compensate for the change of the integral range for the noise part of $\hat{B}(\tau_n)$ [see the form of the corresponding noise component in (\ref{linear})], part of the summation in (\ref{m2}) can be reduced to $\mathcal{T}e^{i\zeta_2\chi\int_0^t d\tau \hat{B}^\dagger\hat{B}(\tau)}\hat{B}(t)$, the form of the evolved linear coupler mode $\hat{B}(t)$ with a time-ordered exponential as the prefactor, giving the evolved mode in Eq. (\ref{solution}) of the main text. The other evolved mode operator $\hat{b}(t)$ is obtained in the same way.

\vspace{0cm}
\subsection* {Appendix D: Waveguide mode evolution under the non-Hermitian effective Hamiltonian}

\renewcommand{\theequation}{D-\arabic{equation}}
\renewcommand{\thefigure}{D-\arabic{figure}}
\setcounter{equation}{0}
\setcounter{figure}{0}

The evolved waveguide modes under the effective Hamiltonian (\ref{pt}) plus the Kerr nonlinear Hamiltonian can be found as follows. Replacing $U^\dagger_L(t)$ in (\ref{tt}) by $U^{-1}_{PT}(t)=e^{iH_{PT} t}$ of the non-Hermitian Hamiltonian 
in (\ref{pt}), we will also have the similar expansion for the non-unitary action 
\begin{eqnarray}
&&U_{eff}(t)=\mathcal{T}e^{-i\int_0^t d\tau(H_{PT}+H_{NL})}\nonumber\\
&=& U_{PT}(t)
\big\{I-i\int_0^t ds_1 U_{PT}^{-1} (s_1)H_{NL}U_{PT}(s_1)\nonumber\\
&-&\int_0^t ds_1 U_{PT}^{-1} (s_1)H_{NL} U_{PT}(s_1)\int_0^{s_1}ds_2 U_{PT}^{-1} (s_2)H_{NL}\nonumber\\
&\times &U_{PT}(s_2) +\cdots \big\}\nonumber\\
&=& e^{-iH_{PT}t}~\mathcal{T}e^{-i\int_0^t d\tau U^{-1}_{PT}(\tau)H_{NL}U_{PT}(\tau)}.
\label{fac-pt}
\end{eqnarray}
The waveguide modes evolve under the first factor on the last line of the above equation to
\begin{eqnarray}
\hat{A}_0(t)
&=&-ie^{-\lambda t}\frac{\eta_1}{J}(i\frac{J}{\eta_1+\eta_2}\hat{a}+\frac{\eta_2}{\eta_1+\eta_2}\hat{b})\nonumber\\
&+&ie^{\lambda t}\frac{\eta_2}{J}(-i\frac{J}{\eta_1+\eta_2}\hat{a}+\frac{\eta_1}{\eta_1+\eta_2}\hat{b}),\nonumber\\
\hat{B}_0(t)
&=&e^{-\lambda t}(i\frac{J}{\eta_1+\eta_2}\hat{a}+\frac{\eta_2}{\eta_1+\eta_2}\hat{b})\nonumber\\
&+&e^{\lambda t}(-i\frac{J}{\eta_1+\eta_2}\hat{a}+\frac{\eta_1}{\eta_1+\eta_2}\hat{b}),
\label{pt-t}
\end{eqnarray}
where the coefficients $\lambda$, $\eta_1$ and $\eta_2$ are the same as in (\ref{linear}).
The transformed Hamiltonian $U^{-1}_{PT}(\tau)H_{NL}U_{PT}(\tau)$ in the second factor of (\ref{fac-pt}) becomes a non-perturbative term with a real number $\lambda$ when $\kappa>J$.

Since the Hamiltonian $H''_{NL}(t)=U^{-1}_{PT}(t)H_{NL}U_{PT}(t)$ is commutative at the different time after neglecting the exponentially decaying terms in (\ref{pt-t}), the action of the second factor in (\ref{fac-pt}) is found as the expansion
\begin{widetext}
\begin{eqnarray}
&&e^{i\int_0^t d\tau U^{-1}_{PT}(\tau)H_{NL}U_{PT}(\tau)}\hat{A}_0(t)e^{-i\int_0^t d\tau U^{-1}_{PT}(\tau)H_{NL}U_{PT}(\tau)}\nonumber\\
&=&\hat{A}_0(t)+[i\int_0^t dt_1 H''_{NL}(t_1),\hat{A}_0(t)]
+\frac{1}{2!}[i\int_0^t dt_2 H''_{NL}(t_2),[i\int_0^t dt_1 H''_{NL}(t_1),\hat{A}_0(t)]]+\cdots\nonumber\\
&=& \hat{A}_0(t)+\sum_{n=1}^\infty\frac{1}{n!}(-i)^n \int_0^t dt_1\chi \tilde{c}_{ab}(t,t_1)\hat{B}_0^\dagger(t_1)\hat{B}_0(t_1)\cdots \int_0^t dt_n \tilde{c}_{bb}(t_{n-1},t_{n})\chi\hat{B}_0^\dagger(t_n)\hat{B}_0(t_n)\hat{B}_0(t_n),
\label{expansion}
\end{eqnarray}
\end{widetext}
where
\begin{eqnarray}
\tilde{c}_{ab}(t,t')&=&[\hat{A}_0(t),\hat{B}_0^\dagger(t')]= i\frac{\eta_2}{J}\zeta_1e^{\lambda t+\lambda t'}
\label{terms-2}
\end{eqnarray}
is the commutator of the linear coupler modes without quantum noise effects, 
and $\tilde{c}_{ab}(t,t')=i\frac{\eta_2}{J}\tilde{c}_{bb}(t,t')$.
With the above results one will obtain the evolved mode
\begin{eqnarray}
&&U_{eff}^{-1}(t)\hat{a}U_{eff}(t)
=i \frac{\eta_2}{J}\hat{B}_0(t)\nonumber\\
&+& i\frac{\eta_2}{J}\sum_{n=1}^\infty \frac{(-i)^n}{n!}\zeta_1^{n}\big(\chi\int_0^t d\tau 
e^{2\lambda\tau}\hat{B}_0^\dagger\hat{B}_0(\tau)\big)^{n} \hat{B}_0(t)\nonumber\\
&=& i \frac{\eta_2}{J} e^{-i\chi\zeta_1\int_0^t e^{2\lambda\tau}\hat{B}_0^\dagger\hat{B}_0(\tau)}\hat{B}_0(t) 
\end{eqnarray}
under the action $U_{eff}(t)$ of the non-Hermitian Hamiltonian. 
The absence of the quantum noises leads to a totally different solution to the evolved waveguide modes.

\vspace{0cm}
\subsection* {Appendix E: Generalization to the multi-mode light fields}

\renewcommand{\theequation}{E-\arabic{equation}}
\renewcommand{\thefigure}{E-\arabic{figure}}
\setcounter{equation}{0}
\setcounter{figure}{0}

In this appendix we present a brief discussion about the generalization to multi-mode light fields. Without loss of generality we consider the light fields in one dimension, i.e, the transverse sizes of the waveguides are negligible. The waveguide modes are then generalized to the multi-mode field operators with the correspondences 
$\hat{a}\rightarrow \hat{\Psi}_a(z,t)$ and 
$\hat{b}\rightarrow \hat{\Psi}_b(z,t)$,
which satisfy the equal time commutation relation $[\hat{\Psi}_c(z),\hat{\Psi}_c^\dagger(z')]=\delta(z-z')$ for $c=a,b$. Here we only consider the Kerr nonlinearity of non-instantaneous action discussed in \cite{bv}, and the quantum field theoretic approach to the instantaneously acting Kerr nonlinearity is similar to those in \cite{mc-4, mc-5, mc-6, mc-8}.

Following the notations in \cite{bv}, we have the dynamical equations corresponding to Eq. (\ref{dynamics}) as follows:
\begin{widetext}
\begin{eqnarray}
&&i\frac{\partial}{\partial z} \hat{\Psi}_a(z,t)=i\kappa\hat{\Psi}_a(z,t)+J\hat{\Psi}_b(z,t)+i\sqrt{2\kappa}\hat{\xi}^\dagger_a(z,t),\nonumber\\
&& i\frac{\partial}{\partial z} \hat{\Psi}_b(z,t)=-i\kappa\hat{\Psi}_b(z,t)+J\hat{\Psi}_a(z,t)+\chi \int d\tau h(t-\tau)\hat{\Psi}_b^\dagger(z,\tau)\hat{\Psi}_b(z,\tau)\hat{\Psi}_b(z,t)+\hat{m}(z,t)\hat{\Psi}_b(z,t)
+i\sqrt{2\kappa}\hat{\xi}_b(z,t).\nonumber\\
\label{multi}
\end{eqnarray}
\end{widetext}
Here the non-instantaneous Kerr nonlinearity with the response function $h(t-\tau)$ necessitates the introduction of the phase noise operator $\hat{m}(z,t)$ for preserving the proper quantum commutation relations. For this type of problem involving a non-instantaneous response function $h(t-\tau)$ it is convenient to replace the evolution along the time axis with that in the wave propagation direction along the $z$ axis \cite{bv}, i.e, to see how the continuous-time field operator $\hat{\Psi}_c(0,t)$ at a starting point on the waveguides evolves to the light field operator $\hat{\Psi}_c(L,t)$ of a certain distance $L$ from the origin. Then, in the above dynamical equations containing the derivative with respect to $z$, one has the corresponding conversion of the dimension for the parameters $\kappa, J$ and $\chi$ by absorbing the other quantities such as the group velocity $v_g$ in them. The evolution operator along the $z$ axis for the process should be a path ordered exponential, which is defined in a similar way to a time ordered exponential. In this practice it is also more convenient to work with the field operator $\hat{\Psi}_c(\tau)$ with the co-moving time variable $\tau=t-z/v_g$, which satisfies the continuous-time commutation relation $[\hat{\Psi}_c(\tau),\hat{\Psi}_c^\dagger(\tau')]=\delta(\tau-\tau')$ (see \cite{bv}).

The procedure in the two previous appendices is also applicable to the current situation. 
The linear combination operator $\hat{\Xi}_{1(2)}(z,t)$ inside the evolved field operators under the linear action [assuming $\chi=0$ and neglecting the phase noise in (\ref{multi})] takes the same form as the corresponding one $\hat{o}_{1(2)}$ in Eq. (\ref{l-solution}) of the main text, except that the single-mode operators are replaced by the light field operators 
$\hat{\Psi}_a(z,t)$ and $\hat{\Psi}_b(z,t)$. The exponentials $e^{\pm \lambda t}$ in (\ref{l-solution}) should be changed to 
$e^{\pm k z}$, where the wave vector $k$ is a function of $\lambda$, since the evolution is along the propagation direction now. There are also the corresponding changes for the noise part of the linearly evolved field operators. The field operators in the factorized 
nonlinear action corresponding to (\ref{infinite-0}) are transformed accordingly, so that the similar procedures to those in Eqs. (\ref{m1}) and (\ref{m2}) lead to the completely evolved field operators. The difference here is that the commutators between the linearly evolved field operators should be changed to $c_{ab}(z,z')$ and $c_{bb}(z,z')$ with the spatial variables. 

Throughout the whole above mention procedure similar to that for the single-mode description, the main term in the expectation value $\langle \hat{\Psi}_b(L, t)\rangle$ of an evolved field operator, which corresponds to the main term in Eq. (\ref{average}), becomes
\begin{eqnarray}
&& \big \langle \alpha_t  \big|e^{i\zeta_2\chi\int_0^L dz e^{2 k z}\int d\tau h(t-\tau)\hat{\Xi}_2^\dagger\hat{\Xi}_2(\tau)}e^{k L}\hat{\Xi}_2(t)\big|\alpha_t  \big\rangle\nonumber\\
&&\times e^{i\zeta_2\chi\int_0^L dz \int d\tau h(t-\tau)\sigma(z)} \mbox{Tr}\big\{\rho_m e^{i\int_0^L dz\hat{m}(z,t)}\big\}.~~~~
\label{m-average}
\end{eqnarray}
The multi-mode coherent state $|\alpha_t \rangle$ in the above equation is the action 
\begin{eqnarray}
e^{\int dt'\alpha(t')\hat{\Psi}^\dagger_a(t')-\int dt'\alpha^\ast(t')\hat{\Psi}_a(t')}|0\rangle=\prod_{\omega} |\alpha(\omega)\rangle
\end{eqnarray}
of a multi-mode displacement operator on a joint vacuum state (the corresponding construction with the spatial distribution is 
given in \cite{mc-5}), and also takes the form of infinite product of coherent states $|\alpha(\omega)\rangle$ with 
$\alpha(\omega)$ being the Fourier transform of the function $\alpha(t)$. The last factor in (\ref{m-average}) is the expectation value of the noise induced phase with respect to the state $\rho_m$ of reservoir giving rise to the phase noise, which can be found with the phase noise correlation function in \cite{bv}. The light field operators in the above equations carry the co-moving time variable. The first factor in (\ref{m-average}) is proportional to the overlap between the transformed multi-mode coherent state $e^{i\zeta_2\chi\int_0^L dz e^{2 k z}\int d\tau h(t-\tau)\hat{\Xi}_2^\dagger\hat{\Xi}_2(\tau)}|\alpha_t \rangle_a|0\rangle_b$ and the input multi-mode coherent state $|\alpha_t  \rangle|0\rangle_b$, and it contains an exponentially accelerating oscillation with respect to the variable $kz$. Hence there are the similar behaviors of the multi-mode mean quadratures and other quantities to those in the single-mode description.


\begin{thebibliography}{99}
\bibitem {book} C. W. Gardiner and P. Zoller, {\it Quantum Noise}, Springer-Verlag, Berlin Heidelberg, 2000.
\bibitem {rv} A. A. Clerk, M. H. Devoret, S. M. Girvin, F. Marquardt, and R. J. Schoelkopf, \rmp 82, 1155 (2010). 
\bibitem {sr} L. Gammaitoni, P. H\"{a}nggi, P. Jung, and F. Marchesoni, Rev. Mod. Phys. 70, 223 (1998).
\bibitem {n-p-t} C. Van den Broeck, J. M. R. Parrondo, and R. Toral, Phys. Rev. Lett. 73, 3395 (1994).
\bibitem {ns} C. Zhou and J. Kurths, Phys. Rev. Lett. 88, 230602 (2002).
\bibitem {y-s} B. Yurke and D. Stoler, Phys. Rev. Lett. 57, 13 (1986).
\bibitem {l-o} P. Kok, W. J. Munro, K. Nemoto, T. C. Ralph, J. P. Dowling, and G. J. Milburn, Rev. Mod. Phys. 79, 135 (2007).
\bibitem{wang} H. Wang, D. Goorskey, and M. Xiao, \prl 87, 073601 (2001).
\bibitem{j1} S. E. Nigg, H. Paik, B. Vlastakis, G. Kirchmair, S. Shankar, L. Frunzio, M. H. Devoret, R. J. Schoelkopf, and S. M. Girvin,
Phys. Rev. Lett. 108, 240502 (2012).
\bibitem{j2} J. Bourassa, F. Beaudoin, J. M. Gambetta, and A. Blais, Phys. Rev. A 86, 013814 (2012).
\bibitem{bender} C. M. Bender and S. Boettcher, Phys. Rev. Lett. 80, 5243 (1998).
\bibitem{bender2} C. M. Bender, Rep. Prog. Phys. 70, 947 (2007).
\bibitem{ex1} A. Guo, G. J. Salamo, D. Duchesne, R. Morandotti, M. Volatier-Ravat, V. Aimez, G. A. Siviloglou, and D. N. Christodoulides, Phys. Rev. Lett. 103, 093902 (2009).
\bibitem{ex2} C. E. R\"{u}ter, K. G. Makris, R. El-Ganainy, D. N. Christodoulides,
M. Segev, and D. Kip, Nature Phys. 6, 192–195 (2010).
\bibitem{ex3}B. Peng, S. K. \"{O}zdemir, F. Lei, F. Monifi,	 M. Gianfreda, G.-L. Long, S. Fan, F. Nori,	C. M. Bender, and L. Yang, Nature Phys. 10, 394 (2014).
\bibitem{ex4} L. Chang,	X. Jiang, S. Hua, C. Yang, J. Wen, L. Jiang, G. Li,	G. Wang, and M. Xiao, Nature Photon. 8, 524 (2014). 
\bibitem{kerr1} R. Tana\'s, ``Nonclassical states of light propagating in Kerr media'', in Theory
of Non-classical States of Light, Editors V. V. Dodonov and V. I. Man'Ko,
Taylor \& Francis, 2003.
\bibitem{jeong} H. Jeong, M. Kang, and H. Kwon, Opt. Commun. 337, 12 (2015).
\bibitem{cat} G. Kirchmair,	B. Vlastakis, Z. Leghtas, S. E. Nigg, H. Paik,	E. Ginossar, M. Mirrahimi, L. Frunzio, S. M. Girvin, and R. J. Schoelkopf, Nature 495, 205–209 (2013).
\bibitem{esd} T. Yu and J. H. Eberly, Science 30, 598 (2009).
\bibitem{n1}H. Ramezani, T. Kottos, R. El-Ganainy, and D. N.
Christodoulides, Phys. Rev. A 82, 043803 (2010).
\bibitem{n2}A. A. Sukhorukov, Z. Y. Xu, and Yu. S. Kivshar, Phys.
Rev. A 82, 043818 (2010).
\bibitem{n3}S. V. Dmitriev, A. A. Sukhorukov, and Yu. S. Kivshar,
Opt. Lett. 35, 2976 (2010).
\bibitem{n5} S. V. Suchkov, B. A. Malomed, S. V. Dmitriev, and Yu. S. Kivshar, Phys. Rev. E 84, 046609 (2011).
\bibitem{n7} D. A. Zezyulin and V. V. Konotop, Phys. Rev. Lett. 108, 213906 (2012).
\bibitem{n8}S. K. Turitsyn, A. M. Rubenchik, M. P. Fedoruk, and E. Tkachenko, Phys. Rev. A 86, 031804(R) (2012).
 \bibitem{n9}A. M. Rubenchik, E. V. Tkachenko, M. P. Fedoruk, and S. K. Turitsyn, Opt. Lett. 38, 4232 (2013).
\bibitem{no1} Z. H. Musslimani, K. G. Makris, R. El-Ganainy, and
D. N. Christodoulides, Phys. Rev. Lett. 100, 030402 (2008).
\bibitem{no2}S. Nixon, L. Ge, and J. Yang, Phys. Rev. A 85, 023822
(2012).
\bibitem{no3}V. Achilleos, P. G. Kevrekidis, D. J. Frantzeskakis, and R.
Carretero-Gonzalez, Phys. Rev. A 86, 013808 (2012).
\bibitem{no4} M. A. Miri, A. B. Aceves, T. Kottos, V. Kovanis, and D. N.
Christodoulides, Phys. Rev. A 86, 033801 (2012).
\bibitem{no5} E.-M. Graefe, J. Phys. A: Math. Gen. 45, 444015 (2012).
\bibitem{no6} Y. Lumer, Y. Plotnik, M. C. Rechtsman, and M. Segev, \prl 111, 263901 (2013).
\bibitem {jing} H. Jing, S. K. \"{O}zdemir, X.-Y. L\"{u}, J. Zhang, L. Yang, and F. Nori, \prl 113, 053604 (2014).
\bibitem{da} D. Dast, D. Haag, H. Cartarius, and G\"{u}nter Wunner, Phys. Rev. A 90, 052120 (2014).
\bibitem{arga} G. S. Agarwal and K. Qu, \pra 85, 031802(R) (2012).
\bibitem{CPT} B. He, L. Yang, Z. Zhang, and M. Xiao, \pra 91, 033830 (2015).
\bibitem {hillery} M. Hillery, Acta Physica Slovaca 59, 1 (2009).
\bibitem {d-w} P. D. Drummond and D. F. Walls, J. Phys. A: Math. Gen. 13, 725 (1980).
\bibitem{num} C. Santori, J. S. Pelc, R. G. Beausoleil, N. Tezak, R. Hamerly, and H. Mabuchi, Phys. Rev. Applied 1, 054005 (2014).
\bibitem{m1} G. J. Milburn and C. A. Holmes, Phys. Rev. Lett. 56, 2237 (1986).
\bibitem{m2} H. Moya-Cessa, Phys. Rep. 432, 1 (2006).
\bibitem{m3} L.-Y. Hu, Z.-L. Duan, X.-X. Xu, and Z.-S. Wang, J. Phys. A: Math. Gen. 44 195304 (2011).
\bibitem{U1} B. He, Phys. Rev. A 85, 063820 (2012).
\bibitem{U2} A. V. Sharypov and B. He, Phys. Rev. A 87, 032323 (2013).
\bibitem{U3} Q. Lin, B. He, R. Ghobadi, and C. Simon, Phys. Rev. A 90, 022309 (2014).
\bibitem{n-h-1} X. Z. Zhang, L. Jin, and Z. Song, Phys. Rev. A 85, 012106
(2012).
\bibitem{n-h-2} M. H. Teimourpour, R. El-Ganainy, A. Eisfeld, A. Szameit, and D. N. Christodoulides, Phys. Rev. A 90, 053817 (2014).
\bibitem {s-v-2} E. V. Shchukin and W. Vogel, \prl 95, 230502 (2005).
\bibitem {s-v-3} W. Vogel and D. G. Welsch, {\it Quantum Optics}, WILEY-VCH, Weinheim, 2006.
\bibitem {s-v-1} E. V. Shchukin and W. Vogel, \pra 72, 043808 (2005).
\bibitem {bs1} M. S. Kim, W. Son, V. Buzek, and P. L. Knight, Phys. Rev. A 65, 032323 (2002).
\bibitem {bs2} X.-B. Wang, Phys. Rev. A 66, 024303 (2002).
\bibitem {mc-1} J. H. Shapiro, Phys. Rev. A 73, 062305 (2006).
\bibitem {mc-2} J. H. Shapiro and M. Rahzavi. New J. Phys. 9, 16 (2007).
\bibitem {mc-3} J. Gea-Banacloche, Phys. Rev. A 81, 043823 (2010).
\bibitem {mc-4} B. He, A. MacRae, Y. Han, A. I. Lvovsky, and C. Simon, Phys. Rev. A 83, 022312 (2011).
\bibitem {mc-5} B. He, Q. Lin, and C. Simon, Phys. Rev. A 83, 053826 (2011).
\bibitem {mc-6} B. He and A. Scherer, Phys. Rev. A 85, 033814 (2012).
\bibitem {mc-7} J. Dove, C. Chudzicki, and J. H. Shapiro, Phys. Rev. A 90, 062314 (2014).
\bibitem {mc-8} B. He, A. V. Sharypov, J. Sheng, C. Simon, and M. Xiao, Phys. Rev. Lett. 112, 133606 (2014).
\bibitem{bv} L. Boivin, F. X. K\"{a}rtner, and H. A. Haus, Phys. Rev. Lett. 73, 240 (1994).
\end{thebibliography}
\end{document}